\documentclass[floats,floatfix,showpacs,amssymb,prd,twocolumn,superscriptaddress,nofootinbib,nolongbibliography,reprint]{revtex4-1}

\usepackage{amssymb,amsmath,verbatim,mathtools,needspace,enumitem,etoolbox,graphicx,physics,microtype,afterpage,xspace,tabularx,lmodern,multirow}
\usepackage{gensymb}
\usepackage[dvipsnames, usenames]{xcolor}
\definecolor{linkcolor}{rgb}{0.0,0.3,0.5}
\usepackage[unicode, colorlinks=true, linkcolor=linkcolor, citecolor=linkcolor, filecolor=linkcolor, urlcolor=linkcolor, linktocpage, breaklinks]{hyperref}
\usepackage[all]{hypcap}
\usepackage[T1]{fontenc}
\usepackage[utf8]{inputenc}
\usepackage[usenames,dvipsnames]{xcolor}
\usepackage[usenames,dvipsnames]{xcolor}
\hypersetup{colorlinks=true,citecolor=romared,linkcolor=romared,urlcolor=romared}

\definecolor{romared}{RGB}{142,0,28}

\newcommand{\be}{\begin{equation}}
\newcommand{\ee}{\end{equation}}

\def\be{\begin{equation}}
\def\ee{\end{equation}}
\newcommand{\beq}{\begin{eqnarray}}
\newcommand{\eeq}{\end{eqnarray}}

\usepackage{aas_macros}
\usepackage{makecell}
\usepackage{soul}
\usepackage[nolist,nohyperlinks]{acronym}

\newcommand{\ba}{\begin{align}}
\newcommand{\ea}{\end{align}}

\newcommand{\Rbh}{R_{\text{a}}}
\newcommand{\Mbh}{M}
\newcommand{\Rc}{R_{\text{c}}}

\newcommand{\altsigma}

\allowdisplaybreaks

\begin{document}

\pagenumbering{arabic}

	%Title of Document	
	\title{Binary superradiance: a numerical study}
	%\title{Superradiance in black hole binaries}
	% Force line breaks with \\
	
	%Authors
  \author{Diogo C. Ribeiro}
	\affiliation{
          Centro de Astrof\'{\i}sica e Gravita\c c\~ao -- CENTRA,
          Departamento de F\'{\i}sica, Instituto Superior T\'ecnico -- IST,
          Universidade de Lisboa -- UL,
          Av.\ Rovisco Pais 1, 1049-001 Lisboa, Portugal
          }

	\author{Miguel Zilhão}
\affiliation{Departamento de Matem\'atica da Universidade de Aveiro and
  Centre for Research and Development in Mathematics and Applications (CIDMA),
  Campus de Santiago,
  3810-183 Aveiro, Portugal}
	\affiliation{
          Centro de Astrof\'{\i}sica e Gravita\c c\~ao -- CENTRA,
          Departamento de F\'{\i}sica, Instituto Superior T\'ecnico -- IST,
          Universidade de Lisboa -- UL,
          Av.\ Rovisco Pais 1, 1049-001 Lisboa, Portugal
          }
	\author{Vitor Cardoso}
	\affiliation{
          Centro de Astrof\'{\i}sica e Gravita\c c\~ao -- CENTRA,
          Departamento de F\'{\i}sica, Instituto Superior T\'ecnico -- IST,
          Universidade de Lisboa -- UL,
          Av.\ Rovisco Pais 1, 1049-001 Lisboa, Portugal
          }
\affiliation{Niels Bohr International Academy, Niels Bohr Institute, Blegdamsvej 17, 2100 Copenhagen, Denmark}

\begin{abstract}
Rotating axisymmetric objects amplify incoming waves by superradiant scattering. When enclosed in a cavity, the repeated interaction of a confined field with the object may trigger superradiant instabilities. Rotating {\it binaries} are ubiquitous in physics, and play a fundamental role in astrophysics and in everyday life instruments.
Such binaries may be prone to superradiant phenomena as well, but their inherent complexity makes it challenging to study how exactly such instabilities can be triggered. Here, we study a binary of two absorbing objects (mimicking black hole binaries, blades of an helicopter, etc) revolving around a common center, and show that superradiant instabilities do occur, on expected timescales and frequency range. Our results provide the first demonstration that superradiance also occurs for highly asymmetric systems, and may have a wealth of applications in fluid dynamics and astrophysics.
Extrapolating to astrophysical black holes, our findings indicate that compact binaries may be used as interesting particle detectors, depositing a fraction of their energy into putative new fundamental ultralight degrees of freedom.
\end{abstract}
	
	%Print Header!
	\maketitle
	
%%%%%%%%%%%%%%%%%%%%%%%%%%%%%%%%%%%%%%%%%%%%%%%
\section{Introduction\label{sec:introduction}}
%%%%%%%%%%%%%%%%%%%%%%%%%%%%%%%%%%%%%%%%%%%%%%%
Energy extraction through superradiance is a fundamental process in physics.
For objects with internal structure, superradiance is a thermodynamic necessity, and follow from the laws of thermodynamics~\cite{ZelDovich1971,ZelDovich1972,Bekenstein:1998nt,Brito:2015oca}.
A rigorous study of superradiance is possible for translational motion, where it can be associated to processes such as the Vavilov-Cherenkov process, the critical speed for superfluidity or superradiance in Mach shocks. These mechanisms all involve superluminal motion~\cite{Bekenstein:1998nt,Brito:2015oca}.

In a pioneering work, Zel'dovich showed that rotating, axisymmetric bodies can also amplify radiation, where now superluminality is replaced by the condition that
the rotational velocity $\Omega$ of the object exceed the rotational velocity $\omega/m$ of the constant-phase surface of the wave~\cite{ZelDovich1971,ZelDovich1972,Brito:2015oca},
\begin{equation}
	\omega < m \Omega \,,	\label{eq:superradiant_condition}
\end{equation}
where we assume a monochromatic wave of frequency $\omega$ and azimuthal number $m$.
Zel'dovich had in mind astrophysical applications, in particular energy extraction from rotating black holes (BHs). The field has since bloomed: BH superradiance is now a well-studied subject ~\cite{Brito:2015oca}, and the investigation of analogue systems has led to the first laboratory measurement of superradiant amplification~\cite{Torres:2016iee}. In the last few years, superradiance from astrophysical BHs has been recognized as an exciting mechanism to probe new fundamental light fields, possibly a component of the elusive dark matter~\cite{Arvanitaki:2010sy,Brito:2013wya,Brito:2014wla,Brito:2017wnc,LIGOScientific:2021jlr,Brito:2015oca}.
Thus, rotational superradiance is now a tool to do particle physics with massive, astrophysical objects.

Axisymmetry plays a key role in our understanding of rotational superradiance, but a plethora of setups of interest are not axially symmetric. This includes astrophysical binaries bound by the gravitational interaction and evolving via gravitational-wave emission, or Earth-bound systems, such as spinning blades encountered in a variety of machinery~\cite{VanBladel:1454380}.
Can superradiant instabilities occur in such non axisymmetric binary systems? If so, on which timescales? 

%%%%%%%%%%%%%%%%%%%%%%%%%%%%%%%%%%%
\section{Setup\label{sec:Setup}	}
%%%%%%%%%%%%%%%%%%%%%%%%%%%%%%%%%%%
To answer these questions, we model the internal degrees of freedom of a binary in a simple yet general way. Being dissipation a key ingredient for superradiance~\cite{Brito:2015oca}, we follow Zel'dovich's work~\cite{ZelDovich1971} and consider the ``dissipative'' Klein-Gordon equation
\begin{equation}
	\square\Psi = \alpha \frac{\partial \Psi}{\partial t} \,, \label{eq:altered_KG}
\end{equation}
to describe the dynamics of a scalar degree of freedom in Minkowsky spacetime. The parameter $\alpha>0$ describes absorption on a time-scale $\tau \sim 1/\alpha$, should the object be at rest in an inertial frame. 

As we shall show, the boundary conditions at the surface of the object implied by such absorption term can be seen to also describe sound waves interacting with cylinders of a given impedance~\cite{Cardoso:2016zvz} if $\alpha \propto Z/( \rho R^3 \omega^2)$, with $\rho$ the density of the medium where a sound wave of frequency $\omega$ propagates, and $Z,R$ the cylinder's impedance and radius respectively. The possibility of manipulating the impedance of a given object~\cite{seddeq2009,quan2014,Yang} motivates and strengthens the analysis below. However, Eq.~\eqref{eq:altered_KG} is intended to mimic more general setups.

In fact, it can also be shown that a sector of electromagnetic waves interacting with a cylinder of a conducting material obeys similar conditions~\cite{Bekenstein:1998nt}.
Model~\eqref{eq:altered_KG} was used as well to model absorption in rotating stars and compact objects~\cite{Cardoso:2015zqa}, where it was shown that in the context of BH physics, $\alpha \sim 1/M$ (with $M$ the BH mass) is the only meaningful choice, and allows to recover known results in BH superradiant scattering~\cite{Cardoso:2015zqa,Brito:2015oca}. 
Recently, the model above was in fact used to study superradiance by binary BHs, within an effective field theory approach, with $\alpha \sim 1/M$~\cite{Wong:2019kru,Wong:2020qom}\footnote{We thank Leong Khim Wong for clarifying this issue for us.}

Outside the ``absorbing'' region, the dynamics of the scalar are simply given by $\square\Psi=0$. For a single spinning cylinder, superradiance was demonstrated by Zel'dovich~\cite{ZelDovich1971}
for the model~\eqref{eq:altered_KG}, and in Refs.~\cite{Cardoso:2016zvz,Bekenstein:1998nt} for sound or electromagnetic waves hitting a cylinder of certain impedance or conductivity, respectively (see the review~\cite{Brito:2015oca}). Here we want to generalize these results to binary systems.

The study of superradiant scattering is challenging because the amplification factors are typically very small. However, the effect can be amplified by placing the system in a cavity, leading to an exponential cascade of energy extraction~\cite{ZelDovich1971,PRESS1972,Cardoso:2004nk,Cardoso:2004hs,Cardoso:2013pza}. This is what we do here.
We take a binary system of two absorbing cylinders revolving around each other at frequency $\Omega$ and at an orbital separation $R_0$. Mathematically, we can model the problem via Eq.~\eqref{eq:altered_KG} by assigning the absorption $\alpha$ the values,
\begin{equation}
	\alpha(t , \mathbf{r}) = 
	\begin{cases}
		\alpha_0 \qquad \text{if} \qquad (\mathbf{r} -\mathbf{R_{orbit})}^2 < \Rbh^2 \\
		\alpha_0 \qquad \text{if} \qquad (\mathbf{r} +\mathbf{ R_{orbit})}^2 < \Rbh^2 \\
		0   \; \;\qquad \text{otherwise}
		\label{eq:alpha_binary_bh}
	\end{cases} \; ,
\end{equation}
where $\mathbf{R_{orbit}}$ defines the orbital radius of the two bodies and $\Rbh=2M$ is taken to be their radius (for definiteness, with a view on astrophysical compact binaries).

Because we eventually would like to extrapolate to gravitational systems, we take the $\Omega$ and the orbital radius to be related through Kepler's law. 
We take the orbital radius to be given by $\mathbf{R_{orbit}} = R_0 \cos (\Omega t)\mathbf{e_x} + R_0 \sin (\Omega t) \mathbf{e_y}$, where $\Omega$ is taken to be the nonrelativistic orbital period
\begin{equation}
\Omega = \sqrt{\frac{2 \Mbh}{R_0^3}} \; .\label{eq:kepler_frequency}
\end{equation}
In the above equation, like throughout our discussion, we have used geometrized units $G = c = 1$.

\begin{figure}[ht]
	\centering
	\includegraphics[width=0.8\linewidth]{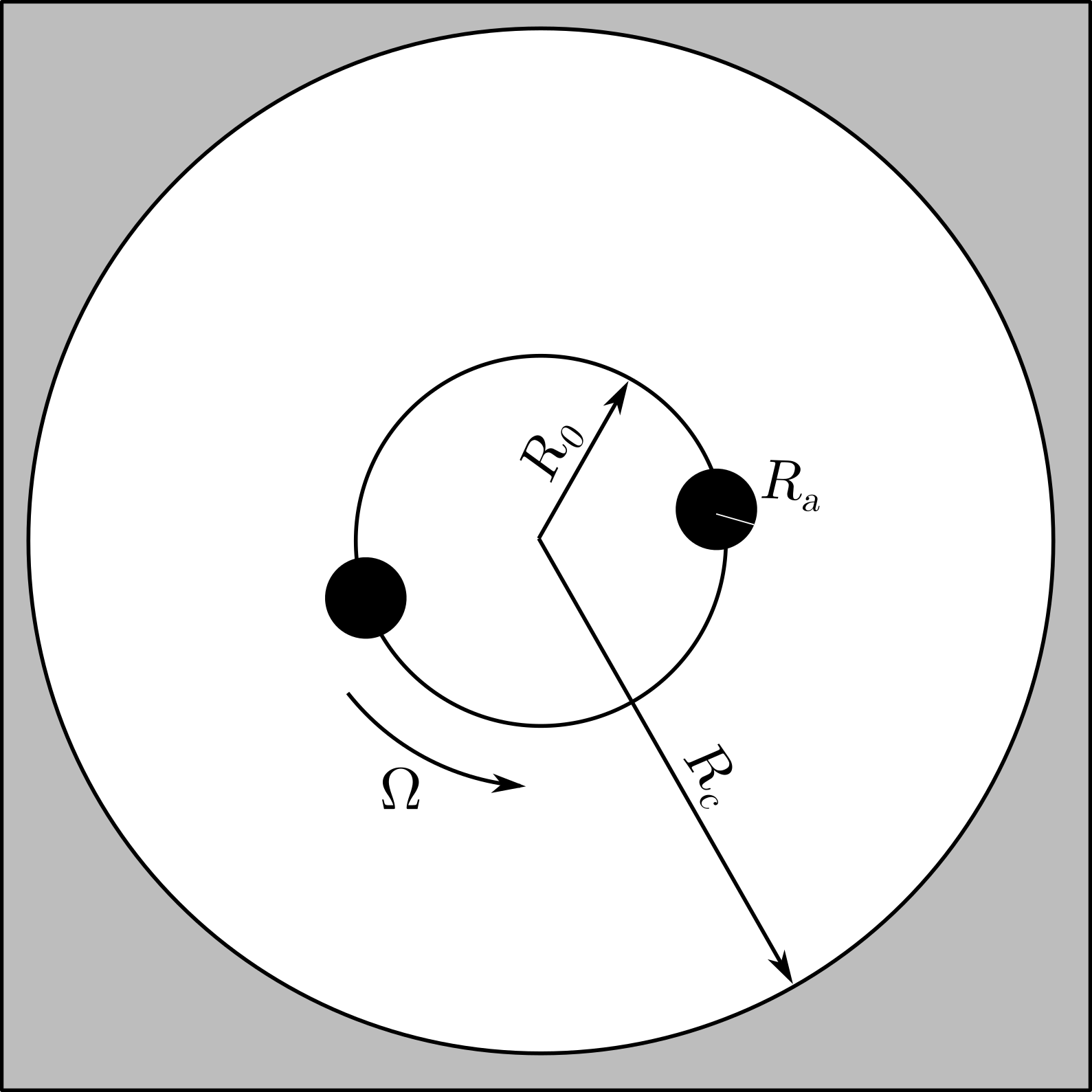}
	\caption{Schematic view of our computational domain, with the respective absorption regions (in black) and reflecting boundary condition at the cavity location, $r = \Rc$. 
		Thus, we consider a binary of equal objects, each with the same radius $\Rbh=2$. They revolve around the geometric center on a circular orbit of radius $R_0$ with frequency $\Omega$ given by~\eqref{eq:kepler_frequency}.
		\label{fig:bhb_squematic}}
\end{figure}

A schematic diagram of our setup is shown in Fig.~\ref{fig:bhb_squematic}. Although we focus on equal mass binaries, the model can easily be generalized.
Note also that previous results concerning a single rotating cylinder can be accommodated setting $R_0=0$ (with $\Omega$ a free parameter).

Even with a simple model as this, it is computationally expensive to explore the problem in $(3+1)$ dimensions so we focus on a $(2+1)$ scenario. 
%\mz{já tínhamos mencionado ser $2+1$ -- ver comentário acima}
Since nothing intrinsic exists about lower dimensional spacetimes, we expect our results to have a counterpart in higher dimensions. We cannot, however, exclude the possibility that the confining geometry here considered plays a role in the dynamics of our system.
Secondly, the generality of the model allows us to draw conclusions about a broader type of systems where the main feature is the clear lack of axial symmetry and thus understand the signature of the system's geometry in a confined field. 

To ease our discussion, we will refer to the single absorption region scenario ($R_0 = 0$) as the SA model while the general binary system ($R_0 \neq 0$) as the BA model.

The governing Eq.~\eqref{eq:altered_KG} was numerically integrated with the aid of \texttt{BhAbs} (Black Hole Absorption Solver) numerical package. This specifically designed code was written in the Julia Programming language~\cite{bezanson2017julia} 
and is freely available in Ref.~\cite{Ribeiro2021}. 

Equation~\eqref{eq:altered_KG} is only valid in the frame where the absorbing regions are static so that one needs to perform a coordinate change to the lab frame. This is easily done through the coordinate change $\varphi \rightarrow \varphi - \Omega t$. Doing so, the governing Eq.~\eqref{eq:altered_KG} can then be written as a set of two first order differential equations
\begin{equation}
\partial_t \Psi = \Pi\,,\qquad \partial_t \Pi  = \nabla^2 \Psi - \alpha \left[ \Pi + \Omega \partial_\varphi \Psi\right] \,,\label{eq:numerical_equations}
\end{equation}
with the boundary condition $\partial_t \Pi= 0$ imposed at the cavity radius $\Rc$.

For the integration of this set of equations, our code implements the method of lines on a Cartesian grid with second-order accurate operators for the discretization of the spatial derivatives, and time integration performed with a fourth order Runge-Kutta scheme provided by the \texttt{DifferentialEquations.jl} Julia package~\cite{Rackauckas2017_DiffEq}.

\begin{table}[t]
	\caption{
		\label{tab:initial_conditions}
		Initial conditions considered for the gaussian pulse~\eqref{eq:initial_pulse}. For the single rotating absorbing region we always used the same initial conditions (model SA). For the binary scenario we considered two sets of parameters whose main difference is the azimuthal mode $m$ (models BA).
			}
	\begin{ruledtabular}
		\begin{tabular}{ l c llll}
			Model & $A$ & $r_0$ & $\sigma$ & $\omega$ & $m$ \\
			\colrule
			SA & 3.5 & 15    & 2.0      & 0.1      & 2 \\
			BA1            & 3.5 & 15    & 2.0      & 0.045    & 2 \\
			BA2            & 5.0 & 35    & 3.5      & 0.045    & 1
		\end{tabular}
	\end{ruledtabular}
\end{table}

For all simulations presented, we take as the initial field configuration a purely ingoing quadrupolar Gaussian wave pulse,
\begin{equation}
\begin{aligned}
& \Psi(t=0,\mathbf{r}) \equiv \Psi_0 = A \cos 2 \varphi \sin \omega r \, e^{ -\frac{1}{2} \left(\frac{r-r0}{\sigma} \right)^2}\,, \\ 
& \Pi(t=0,\mathbf{r}) = \partial_r \Psi_0  \,,
\end{aligned}
\label{eq:initial_pulse}
\end{equation}
where $r_0$, $\sigma$, $\omega$ represent the initial radius, width and frequency of the pulse, respectively. 
For our simulations we considered the three distinct values for these quantities shown in Table~\ref{tab:initial_conditions}.
Due to the linearity of Eq.~\eqref{eq:altered_KG}, the overall amplitude is irrelevant.

The energy density of the field inside the cavity
\begin{equation}
	\epsilon = \frac{1}{V}\int   \left[\left( \partial_t \Psi  \right)^2 + |\nabla \Psi|^2 
	%+ \mu^2|\Psi|^2 
	\right] d \mathbf{x} \; ,
	\label{eq:energy_density}
\end{equation}
was calculated using standard cubic interpolation (Simpson's 3/8 rule). 

The single cylinder 
%\mz{na tabela chamamos-lhe ``single cylinder'', por isso se calhar usar essa nomenclatura sempre? que aliás tabém usamos depois na sec III. verificar se somos consistentes} 
scenario allows us to obtain analytical expressions for the growth rate of a field confined inside the cavity so that a comparison with the numerical simulations is possible. This comparison is presented in Sec.~\ref{sec:single_bh}. The binary case does not gift us with such grace and we restrict ourselves to a purely numerical analysis of the simulations. These results are presented in Sec.~\ref{sec:binary_bh}.

Throughout this manuscript we take 
\be
M = 1\,,\quad \Rbh=2\,,\quad \alpha=10\,.
\label{eq:default_values}
\ee
The first is a choice of scale. The second is chosen with an eye on astrophysical compact binaries (and as we said we will focus exclusively on setups with $\Rbh=2M$) and the third is an arbitrary choice (inspired again by BH physics~\cite{Cardoso:2015zqa}). These are only meant to be representative and to ease the discussion of our numerical results.

%%%%%%%%%%%%%%%%%%%%%%%%%%%%%%%%%%%%%%%%%%%%%%%%%%%%%%%%%%%%%%%%%%%%%%%%%%
\section{Isolated objects: scattering and superradiance \label{sec:single_bh}}
%%%%%%%%%%%%%%%%%%%%%%%%%%%%%%%%%%%%%%%%%%%%%%%%%%%%%%%%%%%%%%%%%%%%%%%%%%
When there is a single spinning cylinder (i.e., $R_0=0$), an analytical solution of Eq.~\eqref{eq:altered_KG} can be obtained in terms of Bessel functions. In polar coordinates $(t,r,\varphi)$, we can use the usual field ansatz
\begin{equation}
	\Psi(t,r,\varphi) = \frac{\phi(r)}{\sqrt{r}} e^{-i \omega t + i m \varphi} \; ,
	\label{eq:decomposition}
\end{equation}
to show that superradiance occurs in this type of system. As Zel'dovich pointed out in~\cite{ZelDovich1971}, performing a Lorentz transformation to the frame at a distance $R_{a}$ from the origin and rotating with velocity $\Omega$, the dissipation term of Eq.~(\ref{eq:altered_KG}) (for the ansatz decomposition above) becomes
\begin{equation}
\alpha \frac{\partial \Psi}{\partial t} \rightarrow i \alpha  \Gamma  \left( \omega - m \Omega\right) \Psi
\end{equation}
where $\Gamma = (1-v^2)^{-1/2}$ is the Lorentz factor and $v =  R_{a} \Omega$ is the instantaneous linear velocity of the frame. When the superradiant condition~\eqref{eq:superradiant_condition} is satisfied, the effective absorption parameter becomes negative, leading to amplification of the field.

Using the same ansatz for the solution and performing the angular coordinate change, the radial component $\phi$ can be seen to satisfy 
\begin{equation}
	\frac{\partial^{2} \phi}{\partial r^{2}}+\left(\omega^{2} + i\alpha (\omega - m \Omega)-\frac{m^{2}}{r^{2}}+\frac{1}{4 r^{2}}\right) \phi=0 \; .
	\label{eq:radial_equation}
\end{equation}
This transformed version of the Bessel equation allows us to write the general solution for the field as
\begin{equation}
	\Psi(t,r,\varphi) = \left[ A  J_m \left( \beta_\alpha r \right) + B Y_m\left(\beta_\alpha r \right) \right] e^{-i \omega t + i m \varphi}\,,
	\label{eq:decomposition_2}
\end{equation}
where $J_m$ and $Y_m$ denote the Bessel functions of the first and second kind, respectively, and $\beta_\alpha^2 = \omega^2 + i \alpha(\omega - m \Omega)$. The whole domain solution consists then of two separate versions of~\eqref{eq:decomposition_2}, one for each region defined that satisfy the appropriate boundary conditions. The general solution must be regular at the origin, be continuously differentiable on the whole domain, and vanish at the cavity radius. This last constraint will force our field to be confined inside the cavity and hence take a particular set of natural frequencies. If, however, we remove this last condition, we are able to study how waves are scattered off the absorbing region.

%%%%%%%%%%%%%%%%%%%%%%%%%%%%%%%%%%
\subsection{Scattering amplitudes}
%%%%%%%%%%%%%%%%%%%%%%%%%%%%%%%%%%

Having no outer boundary means that the field is not confined near the absorbing region. We are thus interested in finding solutions that take the form of a scattering problem,
\begin{equation}
	\Psi(r \rightarrow \infty) \sim \mathcal{A}_+ e^{i \omega r} + \mathcal{A}_- e^{-i \omega r}  \,,
\end{equation}
at spatial infinity, with $\mathcal{A}_\pm$ the amplitude of the outgoing and incoming waves, respectively. To satisfy this condition our solution outside the absorbing region should be written as
\begin{equation}
\Psi(t,r,\varphi) = \left[ \mathcal{A}_+  \phi^+_m \left( \omega r \right) + \mathcal{A}_- \phi_m^- \left(\omega r \right) \right] e^{-i \omega t + i m \varphi}\,,	\label{eq:decomposition_3}
\end{equation}
where $\phi_m^+ = J_m + iY_m$ and $\phi_m^- = J_m - iY_m$ are the Hankel functions of the first and second kind respectively. 

Using the two solutions forms \eqref{eq:decomposition_2} and \eqref{eq:decomposition_3} for the respective regions and requiring continuity at the absorbing region surface we find,
\begin{equation}
	\left|\frac{\mathcal{A}_+}{\mathcal{A}_-}\right|^2 = \left|\frac{(\phi_m^-) (J_m^\alpha)' - (\phi_m^-)' (J_m^\alpha)}{(\phi_m^+) (J_m^\alpha)' - (\phi_m^+)' (J_m^\alpha)}\right|^2 \,,
	\label{eq:scattering_amplitude}
\end{equation}
where primes stand for derivative with respect to the radial coordinate and the functions $J_m^\alpha = J_m(\beta_\alpha r)$ and $\phi_m^{\pm} = \phi_m^{\pm}(\omega r)$ are evaluated at the boundary radius $\Rbh$. 

\begin{figure}[t]
\centering
\includegraphics[width=\linewidth]{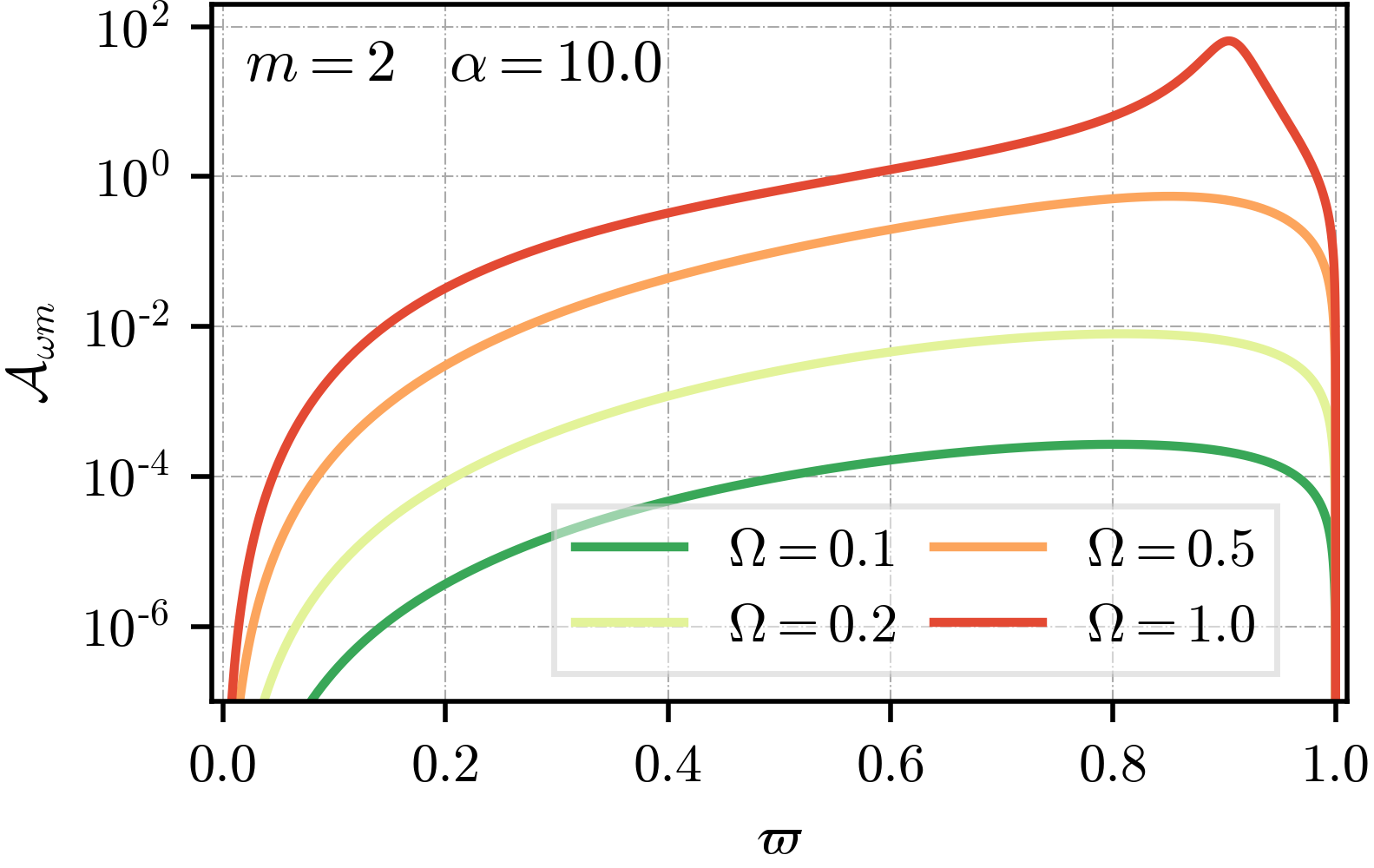}
\caption{Amplification factor $\mathcal{A}_{\omega m} = |\mathcal{A}_+ / \mathcal{A}_-|^2 - 1$ for the SA model with $\alpha=10.0$, as function of the frequency parameter $\varpi = \omega /m \Omega$, for different angular velocity $\Omega$ and an impinging ``quadrupolar'' ($m=2$) wave.	
%The impinging   a ``quadrupolar'' ($m=2$) wave impinging on a rotating cylinder with $\alpha=10.0$, as function of the frequency parameter $\varpi = \omega /m \Omega$, for different angular velocity $\Omega$. 
The radius of the cylinder is $R_{a}=2$, so that some of the configurations are actually superluminal.\label{fig:amplification_factor}
}
\end{figure}
Figure~\ref{fig:amplification_factor} displays the amplification factor $\mathcal{A}_{\omega m} = |\mathcal{A}_+ / \mathcal{A}_-|^2 - 1$ in terms of the parameter $\varpi = \omega / m \Omega$ for a specific set of parameters. For all frequencies below the superradiant condition ($\varpi < 1$) this amplification factor is positive. 
%This behaviour is observed as well in other similar systems~\cite{Cardoso:2016zvz}. 
The frequency at which $\mathcal{A}_{\omega m}$ peaks is usually close to the threshold frequency $m\Omega$, becoming ever so close to this value as the absorption parameter $\alpha$ is increased.  Note that some of the curves correspond to superluminal regimes ($\Omega \Rbh > 1$). 
The behaviour of the scattering amplitudes is very similar to that of sound waves scattering off a uniform cylinder of a given impedance~\cite{Cardoso:2016zvz}. The way the two models are related is discussed in Appendix~\ref{appendix:toy_model_connection}.

%%%%%%%%%%%%%%%%%%%%%%%%%
\subsection{Cavity modes}
%%%%%%%%%%%%%%%%%%%%%%%%%
When the reflecting boundary condition ($\Psi = 0$) is imposed at the cavity radius $\Rc$, our scattering problem turns into an eigenvalue one. Working out the associated algebra allows us to turn the problem of finding the characteristic frequencies into the eigenvalue equation
\begin{equation}
G_m(\omega)=0 \; ,\label{eq:response_function}
\end{equation}
where $G_m$ is a ratio of linear combinations of Bessel functions $J_m$ and $Y_m$ and their derivatives. The actual expression is lengthy and shown in Appendix~\ref{appendix:eigenvalue_equation}. 

For a given set of parameters $(\Rbh, \Rc , \Omega , \alpha , m)$, the roots of $G_m$ correspond to the allowed eigenfrequencies 
\[
\omega=\omega_R+i\omega_I \; ,
\]
that are in general complex-valued. The growing modes are characterized by a positive imaginary part $\omega_I>0$, corresponding to field configurations that grow exponentially in time $\Psi \sim e^{\omega_I t}$. Roots whose imaginary part is negative ($\omega_I<0$) correspond to modes damped in time.

Several other remarks should also be made about the actual dependence of the roots on the chosen parameters. Of main interest to us is the location of the eigenfrequency corresponding to the fastest growing mode. This corresponds to the root with largest imaginary part, ${\rm max} (\omega_I)\equiv \omega_I^{\rm max}$ and will thus dominate the dynamics of the system over large timescales. 
In terms of the absorption parameter, we found that the dependence of $\omega_I^{\rm max}$ with $\alpha$ is roughly linear for small values of $\alpha \sim 0 - 20$. Due to the large size of the parameter space, however, we refrain from attempting to write down a universal law for this behaviour and simply state that this linearity seems to be general in this and similar systems. In particular, this behaviour is also observed for slowly rotating stars~\cite{Cardoso:2015zqa}.

\begin{figure}[t]
\centering
\includegraphics[width=\linewidth]{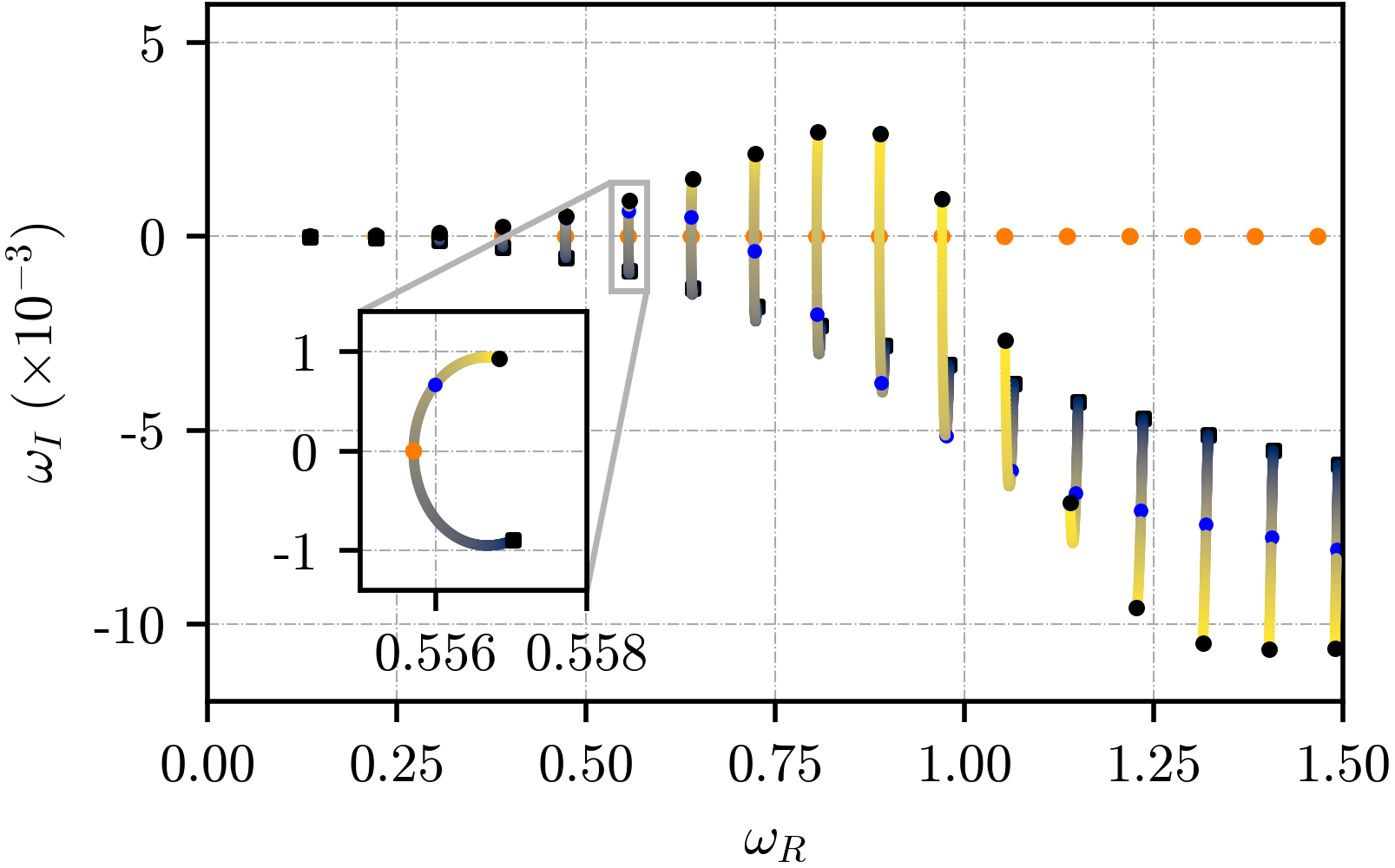}
\caption{Parametric curves of the zeros of the function $G_m(\omega)$ for $\alpha=10.0$, $\Rbh = 2$, $\Rc = 38.0$ and $m = 2$. The black squares indicate the roots for the static case ($\Omega = 0$) while the black circles correspond to the maximally rotating case $\Omega = 0.5$. The lines joining the two are the paths drawn by the zeros as the angular velocity is increased. The orange points on the real axis represent the roots of the Bessel functions of the first kind. The blue points correspond to the roots of a cylinder with angular velocity $\Omega = 0.35$. The inset displays a particular root where the crossing of the real axis can be seen to agree with the Bessel zero. The roots whose real part satisfy the superradiant condition \eqref{eq:superradiant_condition} have positive imaginary part.
\label{fig:parametric_roots_single_bh}
}
\end{figure}

Figure~\ref{fig:parametric_roots_single_bh} depicts how the root structure of $G_m$ depends on the angular velocity of the absorbing region for a particular set of parameters. The lower set of black dots correspond to the natural frequencies of the cavity with the absorbing cylinder at rest, $\Omega = 0$. As expected on physical grounds, all these have negative imaginary parts and hence correspond to damped cavity modes. As the angular velocity is increased, the roots start crossing the $\omega_I = 0$ line and become {\it unstable} modes. The transition occurs precisely when the real part of the roots satisfies the superradiant condition. The top set of black points correspond to the roots of a critically rotating cylinder ($\Omega \Rbh = 1$). 
Similarly to the scattering problem, the maximum instability rate does not correspond to the root whose real part sits closest to the superradiant threshold. Generically this is again controlled by the value of the absorption parameter. Note also the very weak dependence of $\omega_R$ on $\Omega$, as seen in (inset of) Fig.~\ref{fig:parametric_roots_single_bh}.

One important remark is that all azimuthal modes are decoupled. This is readily seen by the fact that our boundary conditions have no azimuthal dependence and hence allow the solutions \eqref{eq:decomposition_2} to evolve independently with the growth rates obtained from the roots structure of the associated $G_m$.  

\begin{figure}[t]
	\centering
	\includegraphics[width=\linewidth]{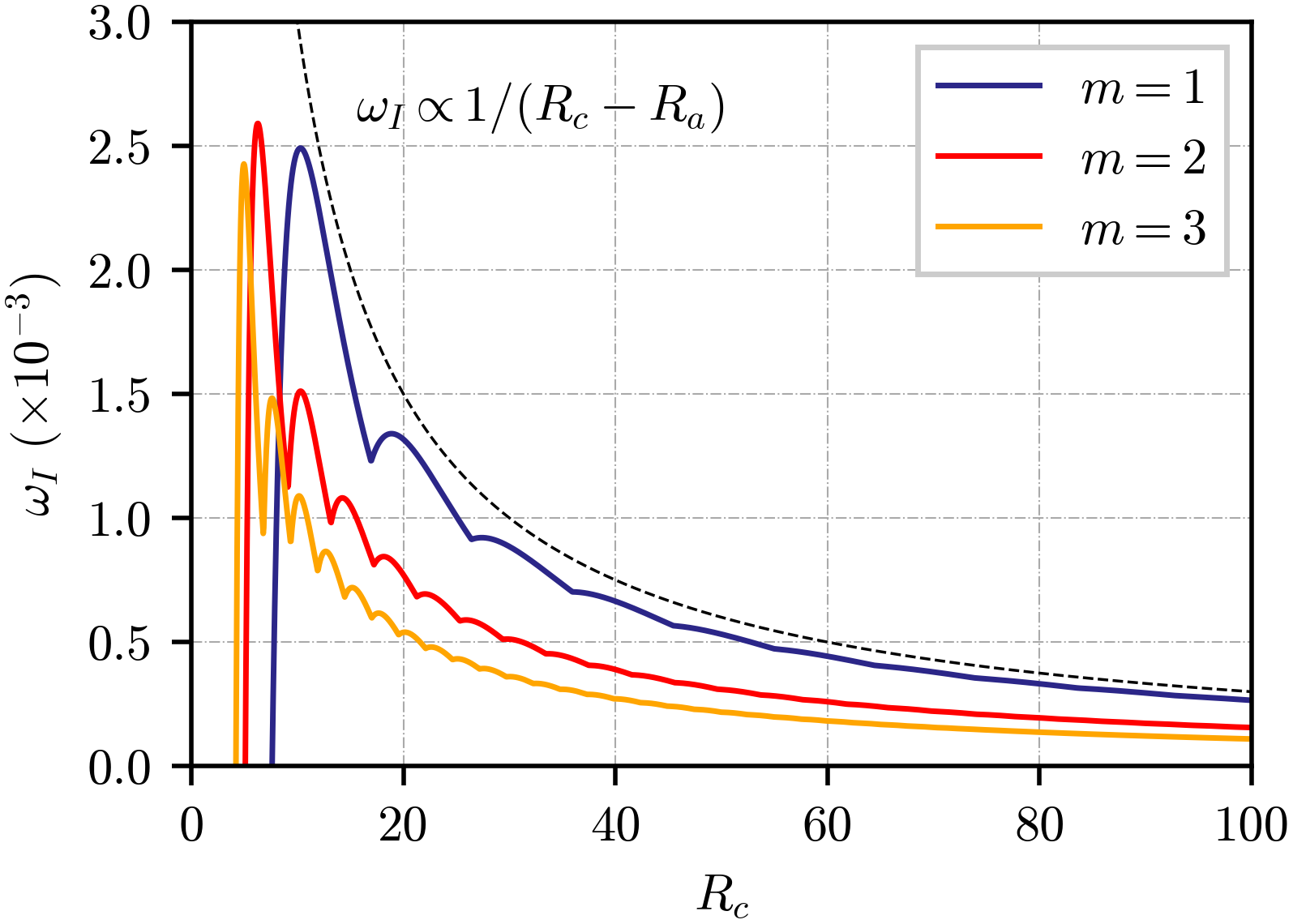}
	\caption{Dependence of the maximum growth rate on the size $\Rc$ of the cavity for the SA model with $\alpha=10,\,\Rbh= 2$ and $\Omega = 0.5$. Each individual concave region of the curve corresponds to a different radial mode. The dashed line depicts the frequency of the traveling pulse inside the cavity. Notice that lower $m$ modes having higher instability rates, at fixed cavity size. For each mode, the threshold is very close to the associated fundamental cavity mode $j_{m,0}/(m \Omega)$.
		\label{fig:amplification_cavity_size_dependence}}
\end{figure}

The dependence of the instability rate on the cavity size is also interesting and is depicted in Fig.~\ref{fig:amplification_cavity_size_dependence} for several azimuthal modes. The dashed line represents the curve proportional to the travel time of a pulse inside the cavity $\tau=(\Rc - \Rbh)^{-1}$. For large cavities, the maximum amplitude seems to follow this behaviour. On physical grounds this is expected since the growth rate of the field should be proportional to how often the pulse interacts with the inner region; nevertheless this overall behavior is interesting, as each individual mode must also be sensitive to the amplifying region itself~\cite{Brito:2015oca}. Interestingly, this behaviour is also observed in the case of an actual BH enclosed in a cavity in (3+1) dimensions but has, to the best of our knowledge, never been pointed out before. We refer the reader to Appendix~\ref{appendix:bh_bomb_growth_rate} for a discussion on this. 

%%%%%%%%%%%%%%%%%%%%%%%%%%%%%%%%%
\subsection{Numerical Comparison}
%%%%%%%%%%%%%%%%%%%%%%%%%%%%%%%%%
%The above is all based on the analytical solution to the problem, in terms of Bessel function. 
Our analytical findings can be corroborated by comparison against the numerical time integration of Eq.~\eqref{eq:numerical_equations}. We consider for this a single absorption region ($R_0=0$) and several values of $\Rc$ and $\Omega$.

\begin{figure*}[tp]
	\centering
	\includegraphics[width=\linewidth]{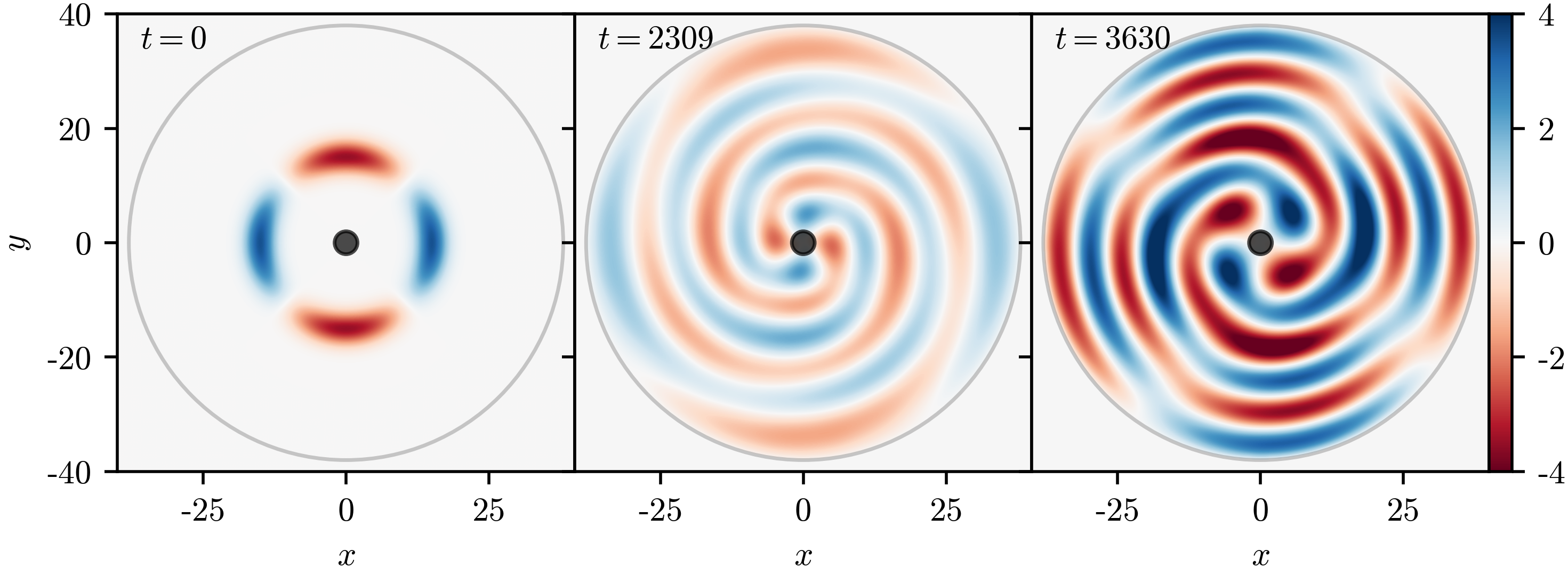}
	\caption{Snapshots of field configuration for three distinct simulation times for an initial Gaussian pulse of the form presented in~\eqref{eq:initial_pulse} with the initial conditions presented in Table~\ref{tab:initial_conditions} (SA model). The cavity has a radius $\Rc=38$ and the absorbing cylinder is rotating with constant angular velocity $\Omega=0.35$. As explained in the text, the angular pattern corresponds to that of a quadrupolar $m=2$ mode, and the rotation of the cylinder excites predominantly a radial mode with overtone $k=5$ ($k=0$ is the fundamental mode), hence with six nodes, as can be seen in the radial profile. }
	\label{fig:single_bh_mural}
\end{figure*}
Figure~\ref{fig:single_bh_mural} features snapshots of the field configuration at three distinct instants of the numerical evolution for a particular simulation where instability of the cavity against superradiance is observed. As time goes by the amplitude of the scalar field increases exponentially. Notice that on the rightmost panel the system evolved for $\sim 200$ revolutions, still a modest number. 

\begin{figure}[thp]
\centering
\includegraphics[width=\linewidth]{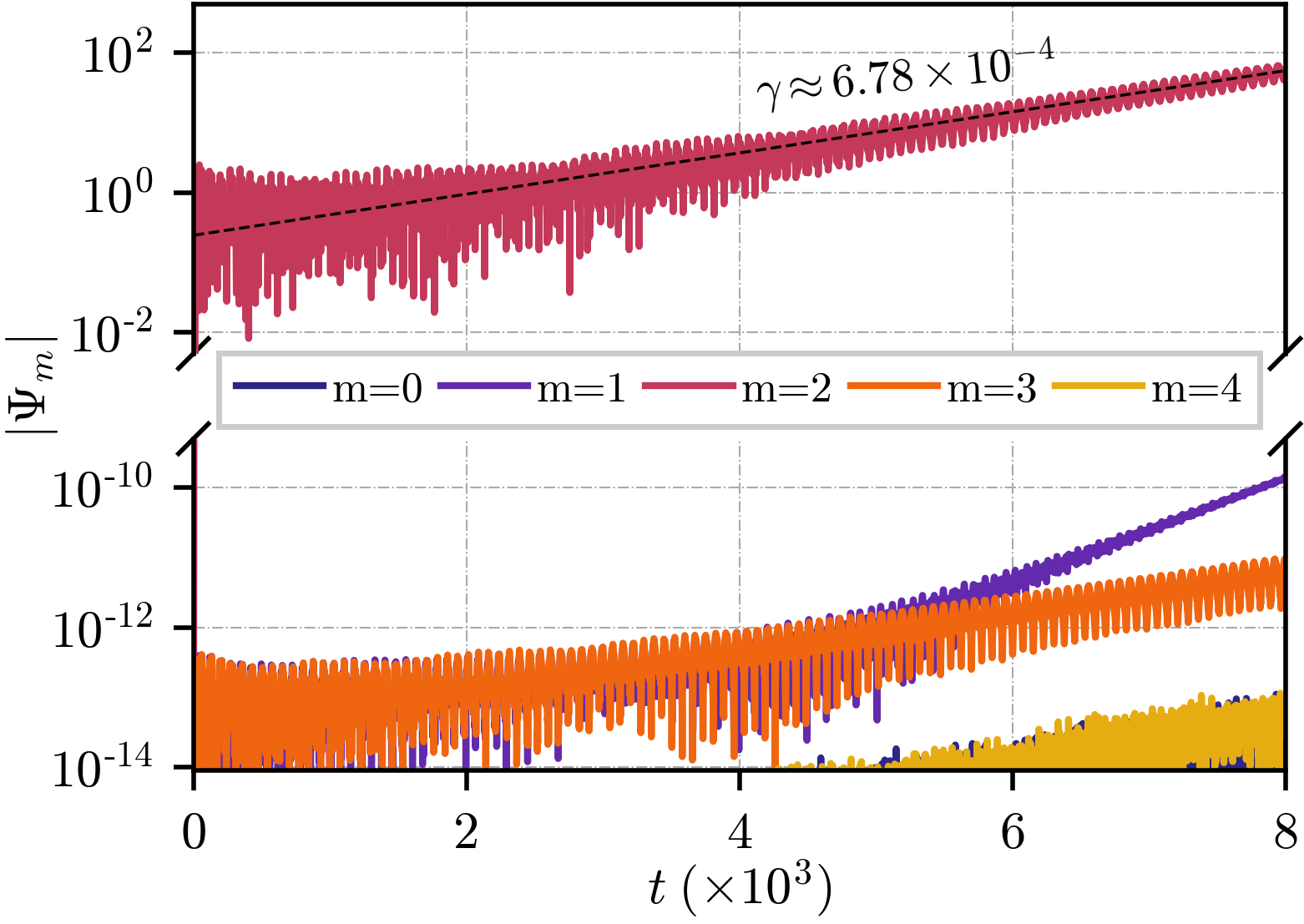}
\caption{
	Mode decomposition of the field at a radius of $r=30$ for an initial Gaussian pulse of the form \eqref{eq:initial_pulse} with the parameters presented in Table \ref{tab:initial_conditions} (SA model)
	in a cavity with radius $\Rc=38$. The absorbing cylinder is rotating with angular velocity $\Omega=0.35$. The dashed black line depicts the analytically obtained growth rate of the fastest growing mode. The numerically obtained frequency of the field is $\omega = 0.5560+0.000677i$. See also Fig.~\ref{fig:single_bh_time_fourier_analysis}.
\label{fig:single_bh_mode_decomposition}}
\end{figure}
The lack of mode mixing mentioned before can be easily confirmed by performing an azimuthal mode decomposition of the field at a given radius,
\[
  \Psi(t,r,\varphi) = \sum_m \Psi_m (t,r) \cos(m \varphi).
\]
The evolution of the Fourier coefficients $\Psi_m$ are shown in Fig.~\ref{fig:single_bh_mode_decomposition} where the field components can be seen to grow exponentially in time, $\Psi_m \sim e^{\gamma t}$. 

The azimuthal mode of the initial pulse is easily seen to dominate the dynamics throughout the simulation with a (numerical) growth rate $\gamma$ that is in accordance with the analytical value for $\omega_I^{\rm max}$ (cf.\ caption in Fig.~\ref{fig:single_bh_mode_decomposition}).

The existence of higher harmonics cannot be completely mitigated due to the Cartesian nature of our numerical grid, but their low amplitude makes the lack of mode mixing evident. The growth rate of these modes was also seen to agree with the analytical values. Finally, and consistently with the results of Fig.~\ref{fig:amplification_cavity_size_dependence}  these lower $m$ modes have larger instability rates and thus -- even when triggered from noise -- will eventually grow to dominate the dynamics.

\begin{figure}[thp]
	\centering
\includegraphics[width=\linewidth]{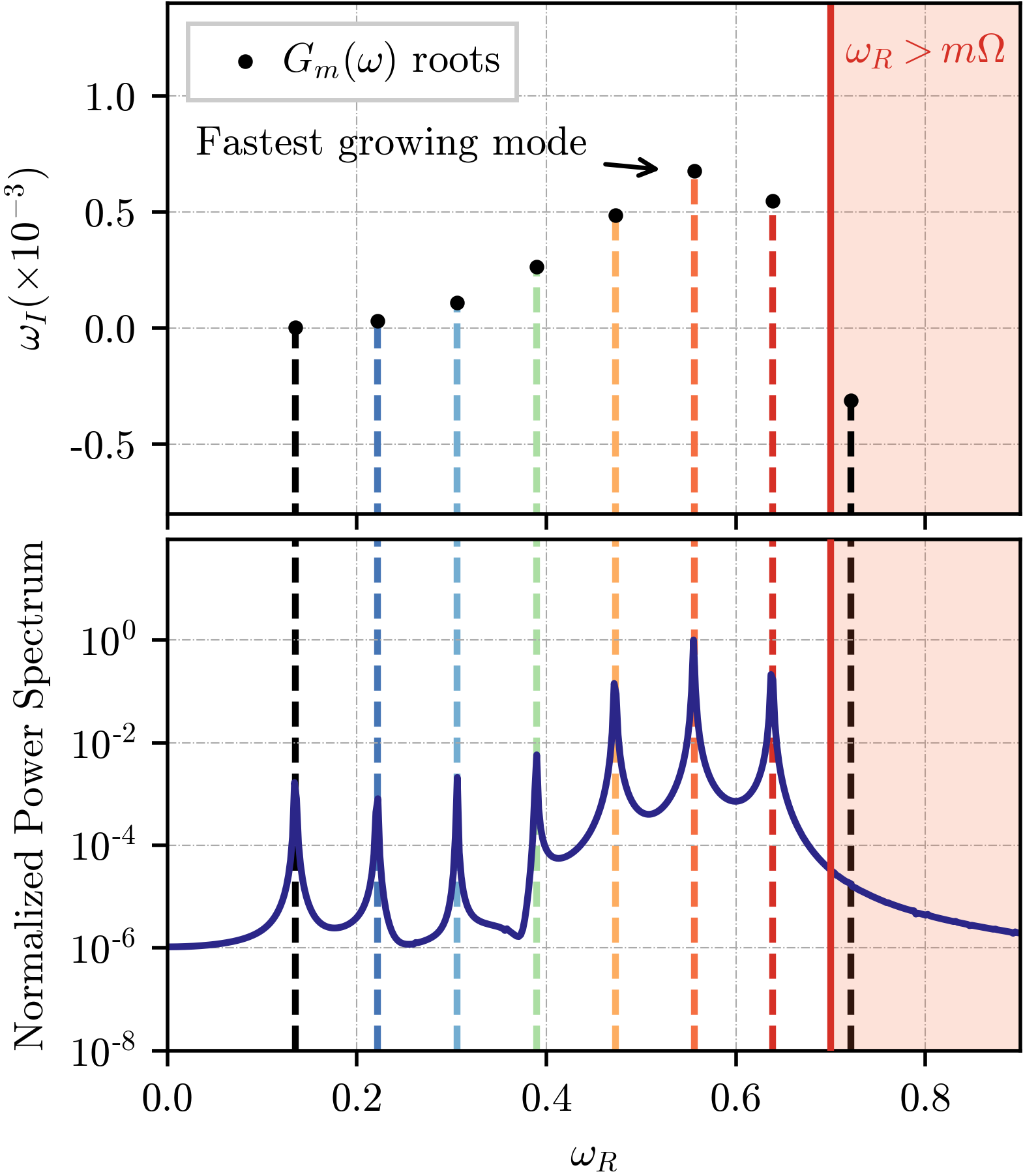}
	\caption{Comparison of the real part of the SA model eigenfrequencies with the numerical data for the simulation whose parameters are presented in Table~\ref{tab:initial_conditions} (SA model).
	The top plot represents the analytical eigenfrequencies in the complex plane. The bottom one depicts the Fourier analysis of the field displacement at $r=20$ for the time interval $\Delta t = [3000, 4000]$. Note that all roots with positive imaginary part lie below the superradiant threshold (red solid line).}
\label{fig:single_bh_time_fourier_analysis}
\end{figure}
The agreement of the real part of the numerically obtained frequency ($\omega_R$) with the analytical one can also be tested by performing a late time Fourier analysis of the field displacement at a fixed point inside the cavity. This analysis is summarized in Fig.~\ref{fig:single_bh_time_fourier_analysis} where one can clearly see the superposition of the several cavity natural frequencies whose imaginary part is positive. Again, note that the fastest growing frequency is not the one closest to the superradiant threshold. In the particular scenario depicted, it corresponds to the 6th Bessel overtone. This can be confirmed visually in Fig.~\ref{fig:single_bh_mural} by counting the number of nodes in the radial direction. 

The growth rate of the energy density of the field inside the cavity, Eq.~\eqref{eq:energy_density}, was also seen to agree with the expected growth rate $\epsilon \sim e^{2\omega_I t}$.

As expected as well, amplification of the field is not always observed~\cite{Cardoso:2004nk}. As already pointed out, the existence of a lower limit on the size of our cavity for which amplification can occur is confirmed by our numerical simulations. The exact value for this size can be obtained from the data of Fig.~\ref{fig:amplification_cavity_size_dependence}. However, a rough estimate can be obtained by noting that the real part of the roots of $G_m$ is very close to the roots of $J_m(\omega \Rc)$, i.e, to the eigenfrequencies of an empty cavity. This implies that the threshold for amplification of the $(m,k)$-th mode is given by
\begin{equation}
	\Rc > \frac{j_{m,k}}{m \Omega} \; ,
\end{equation}
where $j_{m,k}$ is the $k$th zero of the $m$-mode Bessel function of the first kind. For a given $m$, the absolute minimal value of $\Rc$ for amplification to occur is simply $j_{m,0}/(m \Omega)$.

When $\Rc \gg \Rbh$ the excited modes inside the cavity correspond to slightly perturbed empty cavity modes due to the presence of the small absorbing region. This estimate agrees exactly with the results for the amplification of scalar fields by a rotating BH inside a cavity~\cite{Cardoso:2004nk}.

%%%%%%%%%%%%%%%%%%%%%%%%%%%%%%%%%%%%%%%%%%%%%%%%%%%%%%%%%%%%%%%%%
\section{Binaries and cavity resonances\label{sec:binary_bh}	}
%%%%%%%%%%%%%%%%%%%%%%%%%%%%%%%%%%%%%%%%%%%%%%%%%%%%%%%%%%%%%%%%%%
\begin{figure*}[tp]
\centering
\includegraphics[width=\linewidth]{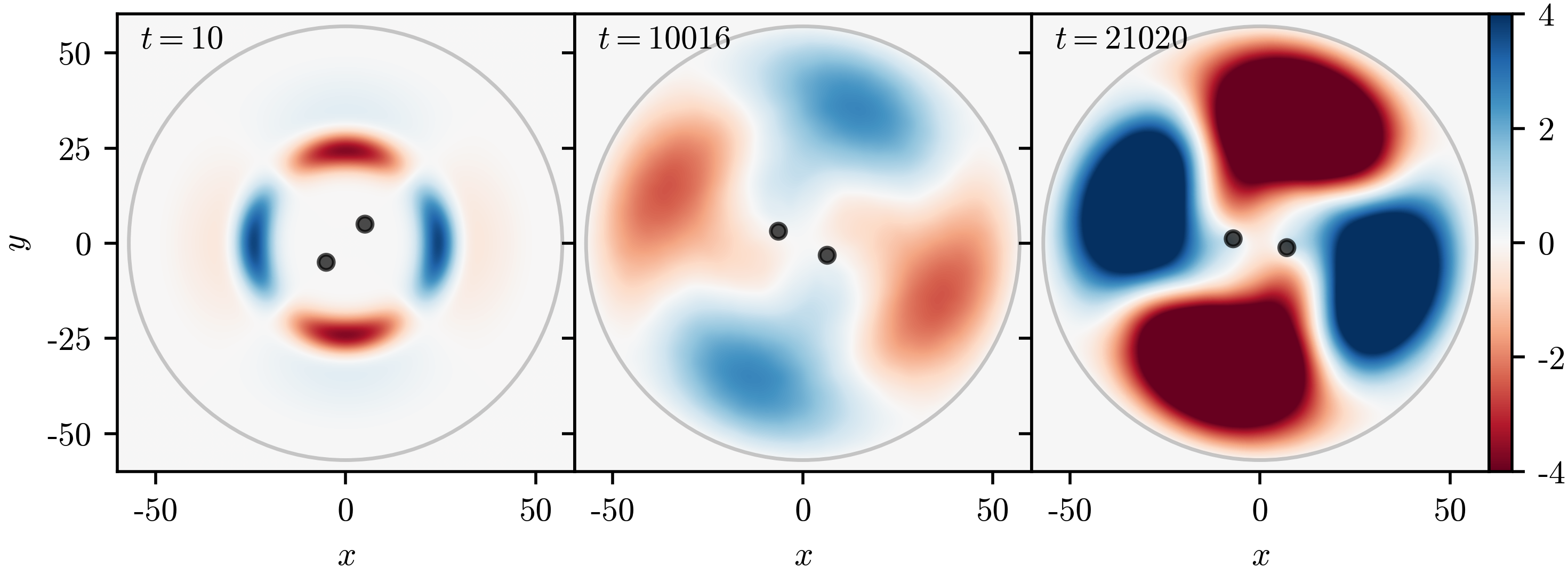}
\caption{Snapshots of field configuration for three distinct simulation times for an initial Gaussian pulse of the form presented in~\eqref{eq:initial_pulse} with the initial conditions presented in Table~\ref{tab:initial_conditions} (BA1). 
	The cavity has $\Rc = 57$ and the two absorbing cylinders are in a circular orbit of radius $R_0 = 7.15$. The orbital period is $T \approx 85$. Note that one could expect frequencies close to $\lesssim m\Omega$ to be excited and hence -- based on Fig.~\ref{fig:single_bh_time_fourier_analysis} -- that the fundamental mode is excited to higher amplitude. This expectaction seems to be supported by the radial profile of the scalar field. See also the main text and Table~\ref{tab:table1}.
\label{fig:binary_bh_mural}
}
\end{figure*}
For the BA model, no analytical expression exists for the eigenfrequencies, and we resort solely to the numerical integration of Eq.~\eqref{eq:numerical_equations} with the appropriate boundary conditions. The numerical convergence of our results is demonstrated in Appendix~\ref{appendix:simulation_convergence}. 

\begin{figure}[tp]
\centering
\includegraphics[width=\linewidth]{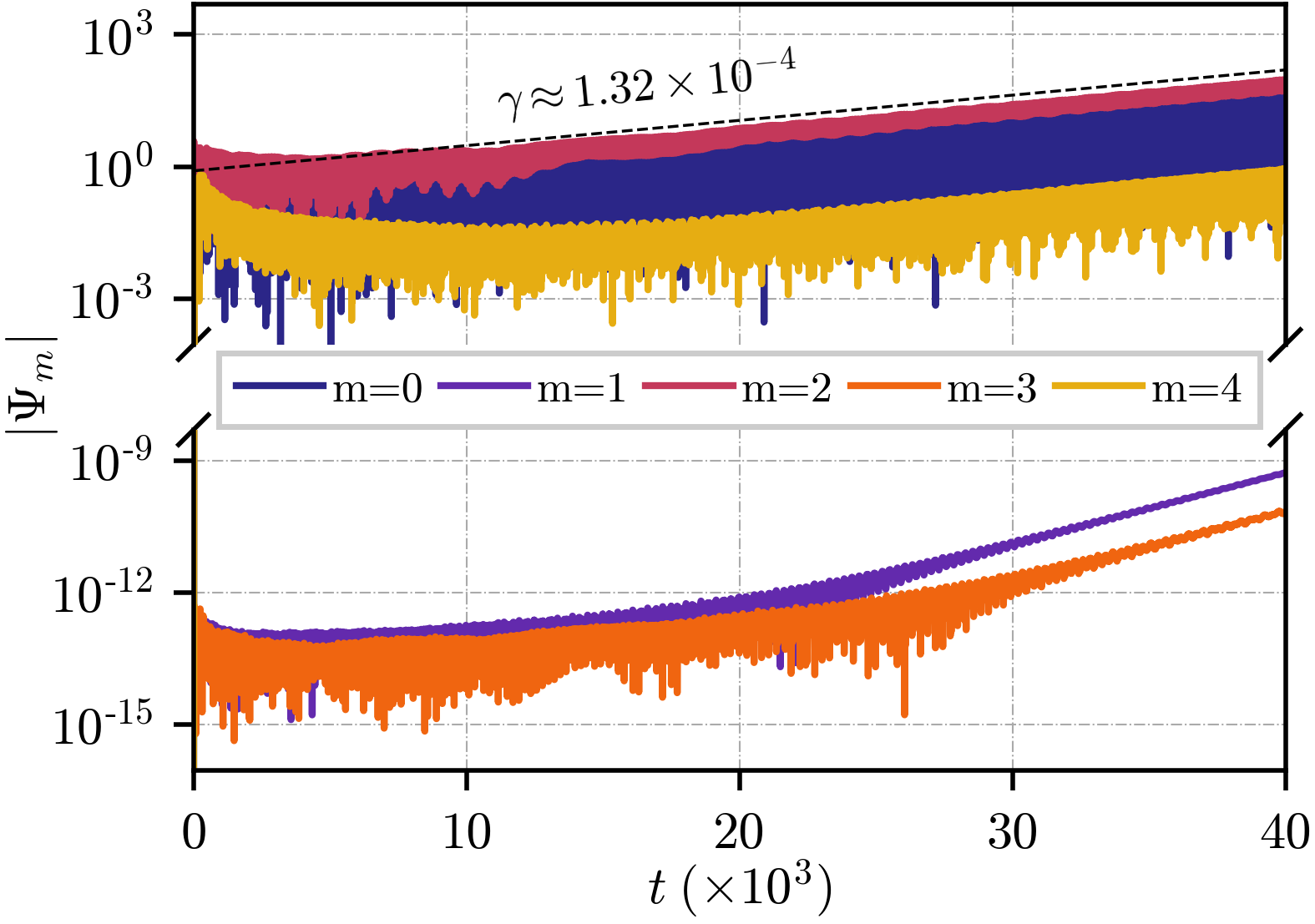}
\caption{Mode decomposition of the field at a radius of $r = 20.0$ for the simulation depicted in Fig.~\ref{fig:binary_bh_mural} (see Table~\ref{tab:initial_conditions} for initial configuration parameters). The initial field azimuthal mode is $m=2$. The generation of higher, even-parity
harmonics is observed from the first interaction with the binary. The dashed black is a linear fit to the $m=2$ curve at late times (shifted upwards). The initial interaction generates other harmonics with the same parity ($m=0,\, 4,\ldots$) that seem to follow the same growth rate. Odd parity modes ($m=1,\, 3,\ldots$) are also present but grow from the numerical noise and have growth rates larger than even modes.
\label{fig:binary_bh_mode_decomposition}}
\end{figure}

\begin{figure}[tp]
\centering
\includegraphics[width=\linewidth]{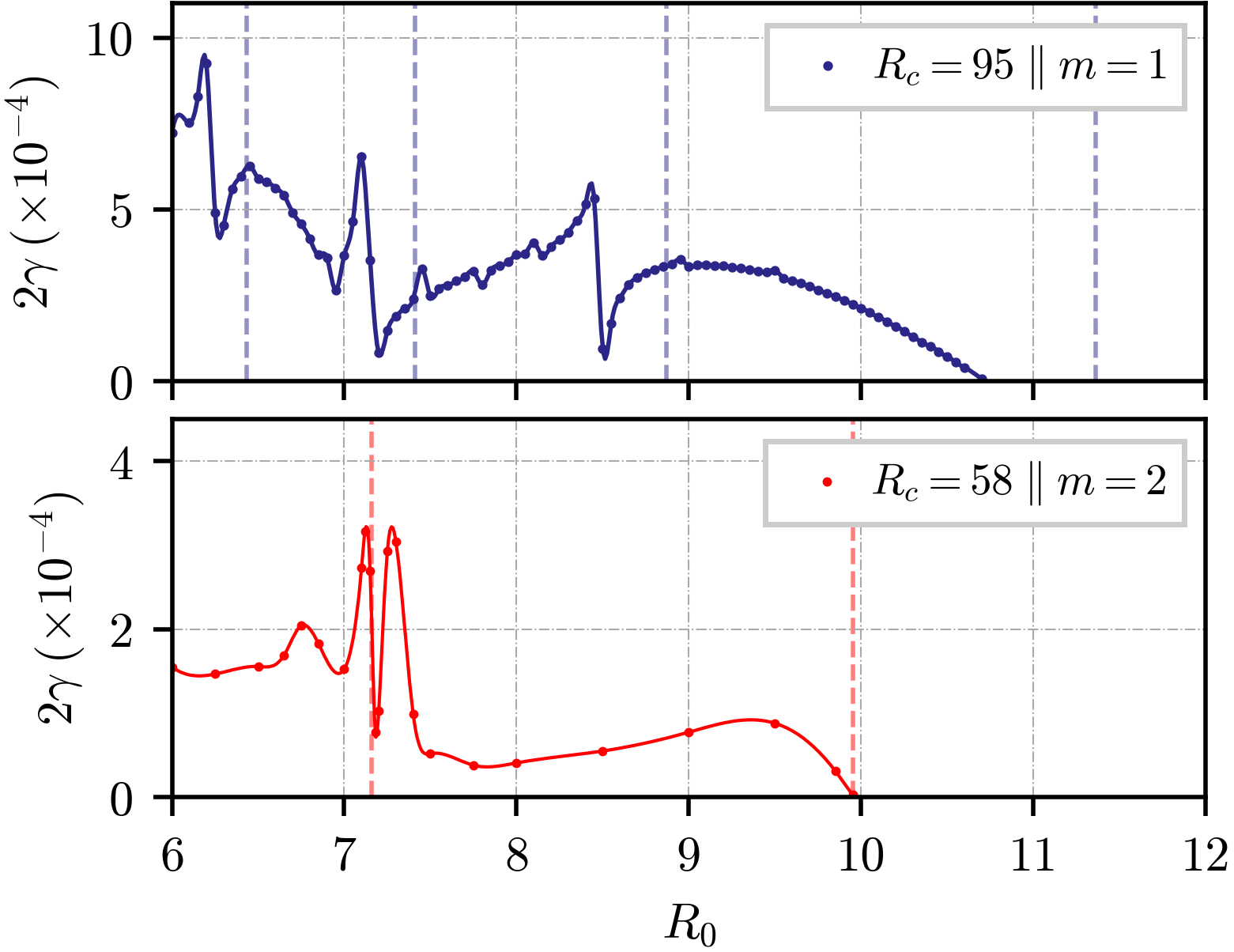}
\caption{Energy growth rate dependence on the orbital radius of the two equal mass binary in a cavity of radius $\Rc=57$ (note the factor two in the axis, since the growth rate should be twice as large as that of the field itself). Dots represent the numerically obtained energy growth rates, while the dashed line corresponds to a quadratic spline. The vertical lines represent the radii defined by Eq.~\eqref{eq:ressonant_radii}. The initial pulse parameters are presented in Table~\ref{tab:initial_conditions}. We used the BA1 (BA2) initial pulse parameters for the bottom (top) plot.
\label{fig:energy_growth_rate_vs_orbital_radius}}
\end{figure}

Our main results are summarized in Figs.~\ref{fig:binary_bh_mural}-\ref{fig:energy_growth_rate_vs_cavity_size}, and are consistent with the excitation of superradiant instabilities in trapped binary systems. This is, to our knowledge, the first solid evidence for such a phenomenon. One example is shown in Fig.~\ref{fig:binary_bh_mural}, where the field configuration inside the cavity is presented for three distinct simulation times. The initial conditions are similar to those used for the SA model in the previous section (see Table \ref{tab:initial_conditions}).

The evolution of the field (its different azimuthal components) at a given radius can be seen in Fig.~\ref{fig:binary_bh_mode_decomposition}. Here as well, we observe amplification of the field. As in the previous scenario, the initial mode of the field ($m=2$) dominates the dynamics of the system. The dashed line in Fig.~\ref{fig:binary_bh_mode_decomposition} corresponds to a linear fit to this mode at late times. The growth rate observed for the specific set of parameters depicted is $\gamma \sim 1.32 \times 10^{-4}$ -- an order of magnitude below the rate for a relatively fast spinning single absorbing region (see Fig.~\ref{fig:single_bh_mode_decomposition}).

To compare the SA model presented in the last section with the binary one here, however, we need to choose an appropriate set of parameters. With the aim of discussing astrophysical scenarios, we consider a single absorption region with $\Rbh = 4$ rotating with the same angular velocity as the binary and inside a cavity of the same size. The comparison of the eigenfrequencies for the two scenarios is shown in Table~\ref{tab:table1}.

The real part of the field frequency $\omega_R$ for both cases is very close. In fact, both configurations excite the first fundamental mode of the cavity. A natural interpretation is that the excitation of a lower energetic cavity mode is mainly caused by a lower frequency of the driving perturber, and not necessarily by a different problem geometry. However, the relation between the radial separation of the binary (a measure of asymmetry) and its orbital velocity \eqref{eq:kepler_frequency} makes this a subtle question to which we shall return at the end of this section.
\begin{table}[t]
\caption{\label{tab:table1}
Comparison of the $m=2$ eigenfrequencies for the BA and SA models (with $R_a=4$ for the latter, intended to describe astrophysical systems with the same total mass) inside a cavity of size $\Rc~=~57$, rotating with the same angular speed $\Omega$. For the binary, $\Omega$ corresponds to $R_0 = 7.15$. The BA eigenfrequency was obtained numerically while the SA is obtained analytically by solving \eqref{eq:response_function}.
}
\begin{ruledtabular}		
\begin{tabular}{ l c c}
  Model  & $\Omega$ & $\omega_R + i \gamma$  \\
  % \hline
\colrule
% \rule{0pt}{3ex}  	
SA	      & 0.07397  & $0.0901 + (8.719 \times 10^{-6})i$               \\
BA        & 0.07397  & $0.0898 + (1.320 \times 10^{-4})i$ 
% \rule{0pt}{3ex}
\end{tabular}
\end{ruledtabular}
\end{table}
The problem's different geometry, nonetheless, seems to be evident when evaluating the growth rate of the modes. From Table~\ref{tab:table1}, the growth rates differ by an order of magnitude. The binary system allows the excitation of the same low energy mode on much shorter timescales. 

Physically, for the same angular velocity, the large separation between the two absorbing regions allows the points within them to be moving much faster, hence allowing a larger angular momentum transfer between the binary and the field. This is the first feature where the geometry of the problem clearly affects the field configuration -- larger growth rates for lower energetic field modes.

The geometry of the system also affects the field dynamics through the coupling of different azimuthal modes. As in the SA model, Fig.~\ref{fig:binary_bh_mode_decomposition} shows us that the initial mode of the pulse dominates the dynamics throughout the simulation. However, its evolution is accompanied by equal-parity modes ($m = 0,4,\ldots$) that grow on similar timescales. 

As before (see Sec.~\ref{sec:single_bh}), the presence of higher odd harmonics cannot be fully mitigated. Despite this, their low amplitude makes it evident that only even modes get coupled to the initial pulse. This type of coupling is present in similarly asymmetric systems (see Appendix~\ref{appendix:appendix_cylinder_model}) and, although we have focused our analysis on initially quadrupolar ($m=2$) field configurations, simulations with an initially $m=1$ pulse also couple with equal parity modes ($m=3,5,\ldots$) and have typically higher growth rates than the even modes.

As mentioned before, the geometrical configuration of the system and the perturbation frequency due to the presence of an absorbing region are intrinsically connected, making the orbital frequency play a crucial but nontrivial role in the long term behaviour of the field's energy content.

To better understand the growth rate dependence on the orbital radius of the binary we performed a large number of simulations sweeping a range of orbital radii. The dependence can be seen in Fig.~\ref{fig:energy_growth_rate_vs_orbital_radius} for two cavity sizes $\Rc = 58$ and $\Rc = 95$. The smaller cavity links with the simulation of Figs.~\ref{fig:binary_bh_mural} and \ref{fig:binary_bh_mode_decomposition} for which an initially Gaussian $m=2$ pulse was used for the field's initial configuration. For the larger cavity, we took instead a pulse with azimuthal number $m = 1$. 

\begin{table}[t]
\caption{\label{tab:table2}
The $m=1$ eigenfrequencies, extracted from late time Fourier analysis for $R_c=95$. Compare the real part of the frequencies with the first two cavity eigenfrequencies $j_{1,0} /R_c =  0.040333$ and $j_{1,1} /R_c =0.738483$. Note that $R_0=7.1, 8.44$ excite the same cavity mode with $j_{1,0} /R_c =  0.040333$.
}
	\begin{ruledtabular}		
		\begin{tabular}{ l c c}
                       $R_0$ & $2\Omega$ & $\omega_{R} + i \gamma$                           \\ %\colrule	\rule{0pt}{3ex}
                        \hline
			6.19  & 0.1837      & $0.07390 + (9.503 \times 10^{-4})i$                     \\ %\rule{0pt}{3ex}
			7.10  & 0.1495      & $0.04034 + (6.531 \times 10^{-4})i$                     \\ %\rule{0pt}{3ex}
			8.44  & 0.1154      & $0.04034 + (5.475 \times 10^{-4})i$ 
                                              % \rule{0pt}{3ex}
                \end{tabular}
	\end{ruledtabular}
\end{table}

The dashed vertical lines correspond to the orbital radii for which the perturbing frequency $2 \Omega$ (note the symmetry of the problem), matches the natural cavity ones, i.e, for which $2\Omega = j_{m,k}/\Rc$. Explicitly we have
\begin{equation}
R_{m,k} = \left[ 2 \left( \frac{2 \Rc}{j_{m,k}}\right)^2 \right]^{1/3}  \; .
\label{eq:ressonant_radii}
\end{equation}
The Keplerian, cavity and scalar frequency are shown in Table~\ref{tab:table2}.

The smaller cavity scenario (bottom panel in Fig.~\ref{fig:energy_growth_rate_vs_orbital_radius}) displays a very clean behaviour over the orbital values considered. As one approaches the resonant $R_{2,1} \approx 7.1$ orbit, the growth rate behaves in a oscillatory manner with overall larger amplification rates. This behaviour is not observed for the second resonant orbital value. Instead, the growth rate plunges towards zero as one approaches it from lower (higher) orbital radii (frequencies). Remarkably, the  $R_{2,1}$ line roughly marks the transition between exciting the $k = 1$ radial mode  (for $R_0 \lesssim R_{2,1}$) and exciting the fundamental $k = 0$ mode (for $  R_{2,1} \lesssim R_0 \lesssim R_{2,0}$). For larger radii, the binary has a lower orbital frequency than the lowest cavity mode and the field never gets amplified.

This seemingly ``clean'' behaviour contrasts with the rather intricate dependence the same value has for a larger cavity (top figure). When $\Rc$ increases, Eq.~\eqref{eq:ressonant_radii} indicates that more resonant orbits are expected to exist in a given interval of orbital radii. Small peaks in growth rate can be seen at the resonant radii, but a large increase in growth rate is not observed at the specified orbits but always at slightly smaller orbits. Also, despite observing the excitation of higher $k$-modes as the orbit is shrunk, no clear transition exists as one crosses the resonances. The fundamental $k=0$ mode dominates the simulation at all timescales for almost all the probed range of radii. The exceptions seem to occur at the resonance orbits where in fact the associated mode seems to be excited. It is also important to note that the rightmost resonance at $R_{1,1} \approx 11.2$ does not correspond to the lowest cavity eigenmode and thus, the growth rate falls below zero much before we reach the limit imposed by \eqref{eq:ressonant_radii}.

The reason for the behaviour above is not entirely clear but two things should be mentioned. First, the width of each peak is much smaller than the radius of the binary cylinders ($\Rbh = 2$). The complex behaviour may be due to the interior freedom of the field in each region. Secondly, the excitation of the fundamental mode ($k = 0$) for the larger cavity points to a nontrivial connection between different $k$-modes (as the fundamental mode may be draining energy from the more energetic ones). This last point may also relate to the geometry of the system, since now different modes are coupled. Even if the frequency of the perturber matches that of the field, the positions of the regions with respect to the cavity mode profile may avoid a proper excitation of the mode.

Finally, perhaps the most important point to retain is the fact that growth rates are large and remain large even when the orbital radius varies by a factor two. This property could be important for astrophysical systems or for Earth-bound experiments.

%%%%%%%%%%%%%%%%%%%%%%%%%%%%%%%%%%%%%%%%%%%%%
\section{Application to black hole physics}
\label{sec:BH_physics}
%%%%%%%%%%%%%%%%%%%%%%%%%%%%%%%%%%%%%%%%%%%%%
In the landscape of General Relativity, BHs are the most well known and visited landmark. They appear as solutions to Einstein field equations and describe vacuum spacetimes with a one-way membrane -- the horizon -- that endows any BH with a natural dissipative mechanism. Besides having very rich phenomenology by themselves, recent interest has sprouted in the area of BH interactions with scalar fields. This interest comes hand in hand with the problem of understanding the nature of the large amount of nonvisible matter we know must permeate our universe~\cite{Roos2010,Bertone2018} -- dark matter.

Although the nature and properties of this exotic kind of matter remain a mystery, ultra-light bosonic fields -- fuzzy dark matter -- have shown to be promising contenders~\cite{Kim1979,Dine1981}. The fuzzy nature of such fields and their weak coupling to the Standard Model makes it extremely hard to probe their properties if not through gravity. BHs, with their extreme gravitational fields, are thus the perfect lab to probe the nature of such Dark Matter candidates.

As previously mentioned, superradiance is intrinsically related to dissipative systems. 
Thus BHs are prone to superradiance. In confining spacetimes, such as asymptotically anti-de Sitter spacetimes, spinning BHs are thus unstable since they behave effectively as BHs in a box, exactly the same setup we studied above~\cite{Cardoso:2004hs,Cardoso:2013pza}.

Spinning BHs may be unstable if new, light bosonic degrees of freedom exist. In this case, BHs would spin-down while growing a bosonic ``cloud'' in their exterior~\cite{Brito:2015oca,Brito:2014wla,Hod2012,Herdeiro2014}. This particularly interesting mechanism allows a rotating BH to transfer energy to the surrounding field provided that condition \eqref{eq:superradiant_condition} is satisfied. The mechanism is similar to the one we studied above, but now confinement is provided by the mass of the bosonic field. The transfer of energy from the rotating BH to the surrounding field is prone to leave clear observational marks~\cite{Arvanitaki:2010sy} that can be used to place strict limits on the mass of ultralight bosons~\cite{Brito:2014wla}.

BH binaries, just like single BHs, present us with an intrinsic dissipation mechanism in the form of an event horizon and so, the natural question arises: are BH binaries, even if composed of nonspinning BHs, prone to similar superradiant phenomena?

The timescales involved together with the nontrivial geometry of a BH binary and radiation losses through gravitational-wave emission make the problem challenging to describe~\cite{Bernard:2019nkv,Ikeda:2020xvt}. Due to the inherent complexity of binary BH spacetimes, the question of if, how and when this phenomenon is relevant in actual astrophysical scenarios remains unanswered.
However, superradiance in BH binary systems was previously shown to occur through an effective field theory approach to the problem~\cite{Wong:2019kru}.

We would like to use our findings above and dwell on BH systems.
We will thus extrapolate our results to BH binaries by promoting $\alpha \sim 1/M$ as has been argued before~\cite{Cardoso:2015zqa}. We should first mention that this substitution yields sensible results: for a single spinning absorbing body, a cavity radius $\Rc=38$ and angular velocity $\Omega \sim 0.5$, the typical amplification scales are of order $\gamma \sim 10^{-3}-10^{-4}$.
This rate is around one order of magnitude larger than the rate of a (3+1) BH bomb with similar cavity radius and BH spin~\cite{PRESS1972,Cardoso:2004nk} with corresponding parameters. 
A thorough comparison of the two models and its eigenfrequencies (as well as the study of the (3+1) equivalent prescription for the absorbing regions) is out of the scope of this work;
Nevertheless, this quick-and-dirty comparison shows that the dissipative model of this work reproduces, within an order of magnitude the correct timescales of a three-dimensional, spinning BH enclosed in a cavity. The instability timescale corresponds to a few seconds for solar mass BHs and to a few months for very massive ones like the one at the center of our galaxy ($M \sim 10^8 M_\odot$).

These results should find a natural application in anti-de Sitter spacetimes. Single spinning BHs were shown to be unstable against superradiant phenomena~\cite{Cardoso:2004hs,Cardoso:2013pza,Chesler:2018txn,Chesler:2021ehz,Brito:2015oca}. Our results indicate that so are binaries, but leave open the nonlinear evolution of such systems.

In an astrophysical context, one should worry about a few issues, one of them being the cavity size and orbital radii.
We did not probe the dynamics of the field for orbital radii smaller than $R_0=6$ since, even without an exact metric, the last stable orbit (LSO) of an equal mass BH binary (BHB) has been evaluated at 2PN order~\cite{Buonanno2000} to be $R_{LSO} \approx 5.718$. After crossing this point, the BHB plunges, making at most a few orbits and then merging into a rotating BH. In this late regime as well, we expect our model to not correctly model the physical system due to the high curvature of the spacetime near the binary. However, the presence of fields can, in principle, affect the late stage dynamics of a plunging BHB.

The LSO value of an equal mass BHB also places a constraint on the smallest possible cavity size for amplification to occur.  In Sec.~\ref{sec:binary_bh} we seeked orbital radii that excited the fundamental modes of a fixed sized cavity. Our result was that of Eq.~\eqref{eq:ressonant_radii}. Turning the question around, we can fix the orbital radius of the BA model and ask what are the cavity sizes for which amplification is enhanced. Like before, we equate the perturbing frequency to the cavity natural ones, $j_{m,k}/\Rc $. We thus obtain
\begin{equation}
\Rc^{m,k}  = \frac{j_{m,k}}{2\Omega} = \frac{j_{m,k}}{2} \sqrt{\frac{R_0^3}{2}} \; .\label{eq:ressonant_cavity_size}
\end{equation}
The above expression represents (for a given orbital radius $R_0$) the threshold cavity size for the amplification of the $(m,k)$-mode. The absolute threshold for the amplification of the fundamental $m$-mode is given by $\Rc > \Rc^{m,0}$.

For a BHB in the LSO, we have $2 \Omega \sim 0.2$ and the crude estimate allows us to say that no amplification is expected for cavities smaller than $\Rc\sim 24.7$. Note that the threshold radius matches exactly the single BH threshold for the $m=2$ case.

\begin{figure}[b]
\centering
\includegraphics[width=\linewidth]{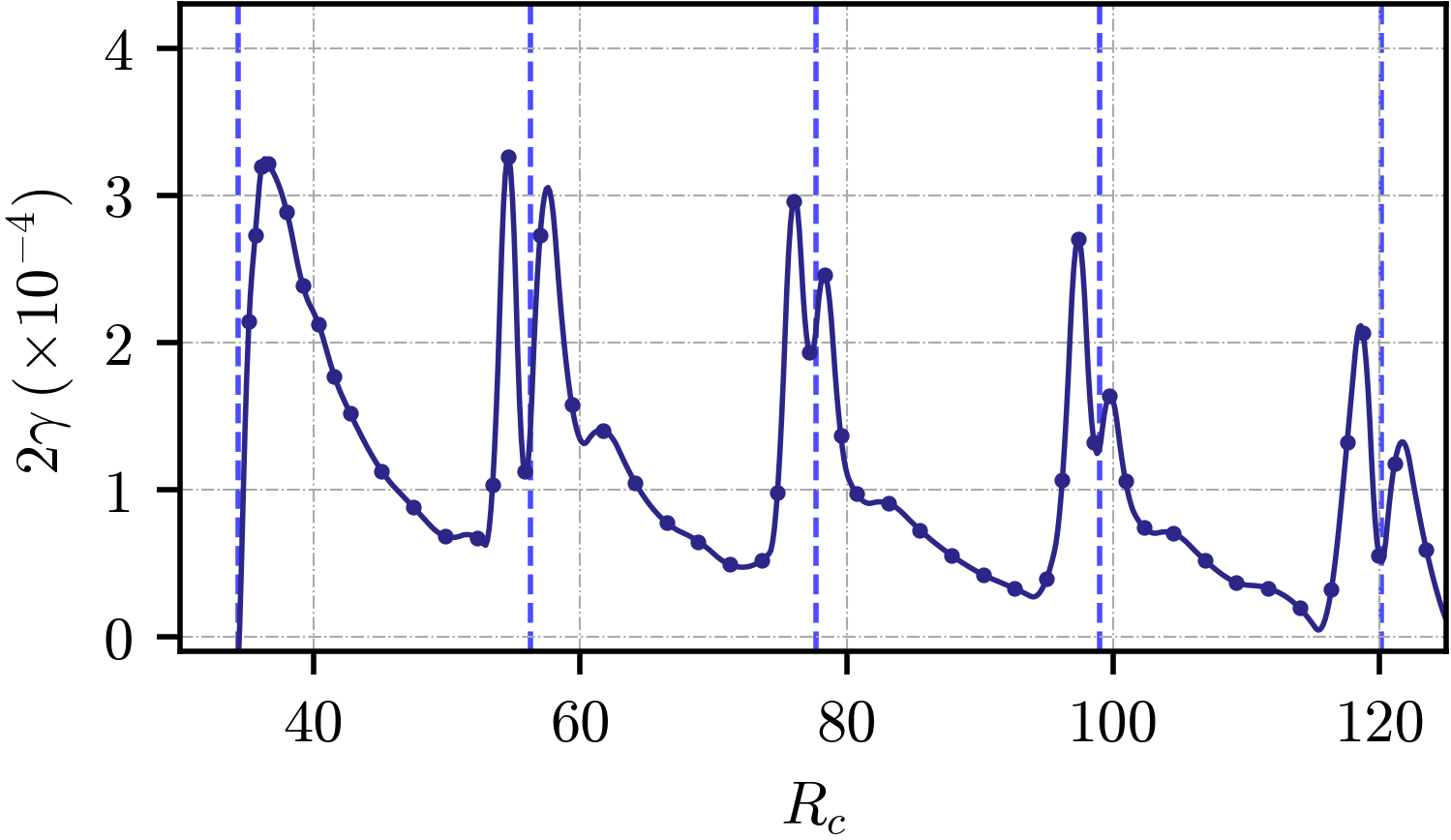}
\caption{Energy growth rate dependence on the cavity radius for the BA with $R_0 = 7.1$. The vertical dashed lines correspond to the resonant cavity \eqref{eq:ressonant_cavity_size}. The initial pulse parameters are given in Table~\ref{tab:initial_conditions}. 
The envelope of the local peaks scale approximately as $\sim 1/\Rc$. The local behavior has a steeper dependence. For example, around the first local peak, we find $\gamma \propto \Rc^{-5}$.  
The initial pulse considered has the BA1 set of parameters presented in Table~\ref{tab:initial_conditions}.
\label{fig:energy_growth_rate_vs_cavity_size} 
}
\end{figure}
Figure~\ref{fig:energy_growth_rate_vs_cavity_size} displays the dependence of the growth rate on the cavity size for a fixed orbital radius of $R_0=7.1$. The dashed lines display the cavity sizes $\Rc^{m,k} $ for the $m = 1$ mode.

The numerically obtained growth rates for the energy field show that in fact, when the frequency of the BHB model matches that of a cavity natural mode, these seem to be larger. It is also evident that our crude estimate of the smaller cavity size for amplification to occur matches the numerics. Note the mild dependence of the instability rate on the cavity size $\Rc$: at large $\Rc$ the local peaks behave as $\sim 1/\Rc$, in agreement with the rate for a spinning BH enclosed in a cavity in (3+1) dimensions (cf. Appendix~\ref{appendix:bh_bomb_growth_rate} and Fig.~\ref{fig:black_hole_bomb}; this fact seems to have gone unnoticed in the literature and requires the study of all the modes of the system).

The observation of superradiant instabilities in such systems is prone to leaving clear observational signatures in both statistical and dynamical studies. In the former class, the loss of energy due to to superradiance may compete with that due to gravitational radiation. For equal mass binaries in a circular orbit, the rate at which the orbital
energy $E_{\rm orb} = M^2/(2 R_0)$ is radiated away is given by~\cite{Peters1963}
\begin{equation}
2 M \gamma_{\rm grav} = \frac{\dot{E}_{\rm grav}}{E_{\rm orb}} =- \frac{64}{10} \left( \frac{M}{R_0}\right)^4  \; .\label{eq:gravitational_wave_energy_loss}
\end{equation}
For a orbit of radius $R_0=8.4$, $2\gamma_{\rm grav} = -1.29 \times 10^{-3}$. For the same orbital radius, the growth rate of a scalar field with azimuthal number $m = 1$ inside a cavity of size $\Rc = 95$  (top panel of Fig.~\ref{fig:energy_growth_rate_vs_orbital_radius}), has an associated value of $2\gamma=5.2 \times 10^{-4} $ -- a rate of comparable magnitude even in the late stages of inspiral. 
%If the results in Fig.~\ref{fig:energy_growth_rate_vs_cavity_size} hold for gravitational systems, with the cavity size $R_c$ proporcional to the inverse Compton wavelength of the field, then at %larger $R_0$ superradiance will dominate over GW emission.

Unlike what Eq.~\eqref{eq:gravitational_wave_energy_loss} implies, the behaviour of the energy loss to superradiance has a more complex dependence on the orbital radius $R_0$ (see Fig.~\ref{fig:energy_growth_rate_vs_orbital_radius}). The existence of radii for which the energy transfer between the binary and the scalar field is more effective, might lead to accelerated plunging of the BHB and give rise to a de-phasing of the GW signals observed when compared with the GW templates. The question of how this signature can be distinguished from other mechanisms
(see, e.g., Refs.~\cite{Cardoso:2011xi,Barausse:2014tra,Annulli:2020lyc}) depends on the actual parameters and must be tackled on an individual basis.

%%%%%%%%%%%%%%%%%%%%%%%%%%%%%%%%%%%%%%%%%%%%%
\section{Discussion\label{sec:discussion}	}
%%%%%%%%%%%%%%%%%%%%%%%%%%%%%%%%%%%%%%%%%%%%%
We provided robust evidence that binaries are also prone to robust superradiant scattering phenomena.
The binary model considered, albeit simple, retains the main geometrical features of many instruments, or laboratory setups. It could describe for example the physics of spinning blades, such as kitchen blenders, or physics associated with helicopter blades. Our model can also describe astrophysical systems, such as BHs or compact stars, and is thus a good starting point to understand the signature left on a bosonic field interacting with such systems.

The two-dimensional model for a single absorbing region, discussed in Sec.~\ref{sec:single_bh} behaves in a similar manner to what is observed in actual $(3+1)$ systems, such as rotating cylinders interacting with sounds waves or electromagnetic waves~\cite{Cardoso:2016zvz,Bekenstein:1998nt}, or even a BH-bomb type scenario~\cite{Cardoso:2004nk}. As pointed out before, the main difficulty in mapping the toy model to the actual scenario lies in choosing a ``correct'' value for the absorption parameter $\alpha$. Our analysis has shown that the amplification factor of the confined scalar field with a single absorbing region depends linearly on the absorption parameter $\alpha$. Note that all our simulations take $\alpha=10$ but the linear behaviour of the growth rate allows us to extrapolate to lower values of this parameter. This linearity was also observed in other scenarios~\cite{Cardoso:2015zqa}, and is consistent with the relation $\alpha\to 1/M$ in BH systems. 

Our main result, however, is the observation of superradiant instabilities triggered by the presence of moving disconnected bodies inside a cavity.
Our result hints at the possibility that superradiant amplification can occur in BH binaries.
The formation and growth of these field configurations can itself radiate gravitationally. The fact that, naturally, BHBs excite lower energetic modes, makes these systems perfect candidates for detection in future detectors aimed at detecting low frequency GW signals \cite{Armano2019}. One of the main scientific objectives of such detectors is the placement of strict constraints in the mass of ultralight bosonic fields. Although we have considered massless scalars, the confining cavity is usually taken as a robust mean of modeling the natural size of scalar clouds around BHs~\cite{Hui2019}. 

The confinement of the field may also arise due to density gradients in the interstellar medium~\cite{Vicente:2019ilr} (but see also Ref.~\cite{Cardoso:2020nst}). In this scenario, the pressure exerted on the cavity walls (the interstellar medium) can play a crucial role in the dynamics of astrophysical objects. The details of such interaction, however, require a better knowledge of the nature of the scalar matter and we refrain from commenting on this any further.

%%%%%%%%%%%%%%%%%%%%%%%%%%%%%%%%%%%%%%%
\begin{acknowledgments}
We are grateful to Leong Khim Wong for providing useful feedback and comments on a version of this manuscript.
V.C.\ is a Villum Investigator supported by VILLUM FONDEN (grant no. 37766) and a DNRF Chair support by the Danish National Research Foundation. 
M.Z.\ acknowledges financial support provided by FCT/Portugal through the IF programme, grant IF/00729/2015, and
by the Center for Research and Development in Mathematics and Applications (CIDMA) through the Portuguese Foundation for Science and Technology (FCT -- Funda\c{c}\~ao para a Ci\^encia e a Tecnologia), references UIDB/04106/2020, UIDP/04106/2020 and the projects PTDC/FIS-AST/3041/2020 and CERN/FIS-PAR/0024/2021.
This project has received funding from the European Union's Horizon 2020 research and innovation programme under the Marie Sklodowska-Curie grant agreement No 690904. 
We thank FCT for financial support through Project No.\ UIDB/00099/2020 and through
grants PTDC/MAT-APL/30043/2017 and PTDC/FISAST/7002/2020.
We further acknowledge support from the European Union's Horizon 2020 research and innovation (RISE) program H2020-MSCA-RISE-2017 Grant No.\ FunFiCO-777740.
We acknowledge that the results of this research have been achieved using the DECI resource Snellius based in The Netherlands at SURF with support from the PRACE aisbl, as well as the ``Baltasar-Sete-Sóis'' cluster at IST.
\end{acknowledgments}
%%%%%%%%%%%%%%%%%%%%%%%%%%%%%%%%%%%%%%
\appendix

%%%%%%%%%%%%%%%%%%%%%%%%%%%%%%%%%%%%%%%%%%%%%%%%%%%%%%%%%%%%%%%%
\section{Connection with cylinder model \label{appendix:toy_model_connection}}
%%%%%%%%%%%%%%%%%%%%%%%%%%%%%%%%%%%%%%%%%%%%%%%%%%%%%%%%%%%%%%%%

The scattering problem treated in Sec.~\ref{sec:single_bh} allows us to bridge the gap between our model and previously studied ones. Namely, our model is in every aspect analogous to that of a cylinder of radius $R$ and a given impedance $Z_0$ immersed in a fluid of density $\rho$ \cite{Cardoso:2016zvz}. The scattering amplitude for such system is given by
\begin{equation}
	\frac{|\mathcal{A}_+|^2}{|\mathcal{A}_-|^2} = \left | \frac{ (1-1/\sigma) \phi^{-}_{m}-i Z\left(\phi^{-}_{m}\right)^{\prime}}{(1-1/\sigma) \phi^{+}_{m}-i Z\left(\phi^{+}_{m}\right)^{\prime}} \right |^2 \; ,
	\label{Appendix:scattering_TM1}
\end{equation}
where $Z = Z_0 / (\rho c)$, $\sigma = \omega / m \Omega$ and the prime denotes differentiation with respect to $y = \omega r / c$. Comparing this expression with Eq.~\eqref{eq:scattering_amplitude} we can conclude that the two models behave similarly as long as
\begin{equation}
	i \left(\frac{Z}{\rho c}\right) \left(\frac{c}{\omega}\right)\left( \frac{\sigma}{\sigma - 1}\right) = \frac{J_m^\alpha}{(J_m^\alpha)'}
\end{equation}
is satisfied. In this expression, the derivative should be taken with respect to $r$ and evaluated at the cylinder radius $R$. By expanding the right-hand-side of the above equation in powers of $(\beta_\alpha r)$ a relation between the models can be obtained. Consider a static cylinder $\sigma/(\sigma -1) \rightarrow 1$. To $\mathcal{O}((\beta_\alpha R)^3)$ order then, the mapping
\begin{equation}
\alpha \rightarrow i \omega + \frac{2m^2(m+1)}{\omega R^3} \left[ \frac{ Z}{\rho \omega} + i\frac{R}{m}\right] \; .
\label{eq:toy_model_connection}
\end{equation}
formally makes the scattering amplitudes from a cylinder of impedance $Z$ equivalent to those from a cylinder of absorption $\alpha$. Since this procedure only depends on the boundary conditions of the cylinder model and the interior solutions of the one presented in Sec.~\ref{sec:single_bh}, the connection~\eqref{eq:toy_model_connection} remains valid when the system is enclosed in a cavity.

%%%%%%%%%%%%%%%%%%%%%%%%%%%%%%%%%%%%%%%%%%%%%%%%%%%%%%%%%%%%%%%%%%%%%%%
\section{Eigenvalue equation\label{appendix:eigenvalue_equation}	}
%%%%%%%%%%%%%%%%%%%%%%%%%%%%%%%%%%%%%%%%%%%%%%%%%%%%%%%%%%%%%%%%%%%%%%%
Function~\eqref{eq:response_function} is obtained by equating the general solution for the field inside the cavity \eqref{eq:decomposition_2} and impose on it the appropriate boundary conditions. To do so, we take two versions of Eq.~\eqref{eq:decomposition_2} (one for the region inside the cylinder and another one outside it) and require: regularity at the origin ($r= 0$), continuity (field and derivative) at the absorbing region interface, vanishing boundary condition at the cavity radius. Doing the algebra, we see that $G_m$ is given by
\begin{widetext}
	\begin{equation}
		G_m(\omega) = \frac{ J_m \left( \beta_0 \Rbh \right)}{J_m \left( \beta_\alpha \Rbh \right)} + \left(\frac{D}{C} \right) \frac{ Y_m \left( \beta_0 \Rbh \right)}{J_m \left( \beta_\alpha \Rbh \right)} -  \frac{ J_m' \left( \beta_0 \Rbh \right)}{J_m' \left( \beta_\alpha \Rbh \right)} - \left(\frac{D}{C} \right) \frac{ Y_m' \left( \beta_0 \Rbh \right)}{J_m' \left( \beta_\alpha \Rbh \right)}\,,
		\label{eq:full_function}
	\end{equation} 
\end{widetext}
where $D/C$ is given by
\begin{equation}
	\frac{D}{C} = -\frac{J_m(\beta_0 \Rc)}{Y_m(\beta_0 \Rc)}\,,\label{eq:DC_ratio}
\end{equation}
and the derivatives are taken with respect to the radial coordinate $r$. The roots of~\eqref{eq:full_function} are obtained by usual root finding procedures.

%%%%%%%%%%%%%%%%%%%%%%%%%%%%%%%%%%%%%%%%%%%%%%%%%%%%%%%%%%%%%%
\section{Black hole bomb \label{appendix:bh_bomb_growth_rate}}
%%%%%%%%%%%%%%%%%%%%%%%%%%%%%%%%%%%%%%%%%%%%%%%%%%%%%%%%%%%%%%

\begin{figure}[ht]
	\centering
	\includegraphics[width=\linewidth]{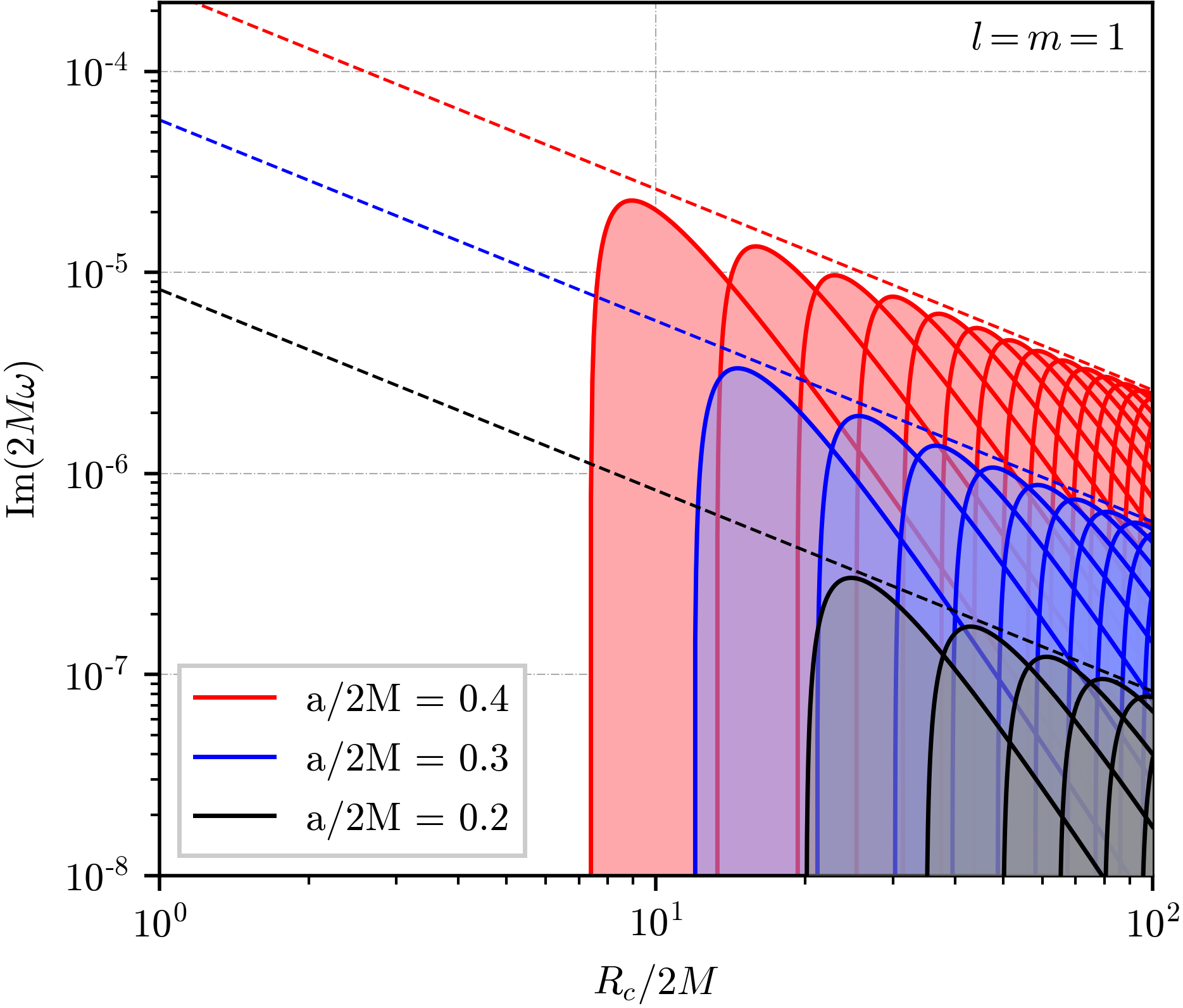}
	\caption{Imaginary part of the eigenfrequencies BH bomb as a function of the cavity radius $\Rc$ for several values of the rotation parameter $a$ and $l = m = 1$. Each solid line corresponds to a different cavity radial mode ($n = 1,2,3,\ldots$). Note that there for each $n$ there exists a minimal cavity radius for which the mode can be amplified. \label{fig:black_hole_bomb}
	}
\end{figure}

The confinement of a scalar field near a rotating BH was first studied by Press and Teukolsky~\cite{PRESS1972}. The BH bomb, as the system was called therein, was later seen to develop instabilities~\cite{Cardoso:2004hs}. Just like in our model, the field inside the cavity is characterized by having a specific eigenfrequency with a real and imaginary part. These eigenfrequencies are found by solving Teukolsky canonical equation \cite{Cardoso:2004hs}
\begin{equation}
	\frac{d^2 Y}{d r_\star^2} + \left[\frac{K^2 - \lambda \Delta}{(r^2 + a^2)^2} - G^2 - \frac{dG}{d r_*} \right] Y = 0 \; ,
	\label{eq:Teukolsky_equation}
\end{equation}
where $r_*$ is the tortoise coordinate. The equation is easily solved by a shooting method, starting from the BH horizon up to the cavity radius. The eigenfrequencies obtained trough this method have been extensively studied \cite{Cardoso:2004hs}. They have a similar behaviour to the individual roots seen for our model and the one in \cite{Cardoso:2016zvz}. However, the inverse behaviour with the cavity radius pointed out in Fig.~\ref{fig:energy_growth_rate_vs_cavity_size} was not previously observed in the BH bomb scenario. Figure~\ref{fig:black_hole_bomb} depicts this behaviour for the $l=m=1$ and several BH rotation speeds. This global behavior seems to agree with the naive expectation that the growth rate of the field should be proportional to the inverse interaction time of a traveling pulse and the event horizon.

%%%%%%%%%%%%%%%%%%%%%%%%%%%%%%%%%%%%%%%%%%%%%%%%%%%%%%%%%%%%%%%%%%%%%
\section{Convergence analysis\label{appendix:simulation_convergence}}
%%%%%%%%%%%%%%%%%%%%%%%%%%%%%%%%%%%%%%%%%%%%%%%%%%%%%%%%%%%%%%%%%%%%%
%
\begin{figure*}[th]
\centering
\includegraphics[width=0.475\linewidth]{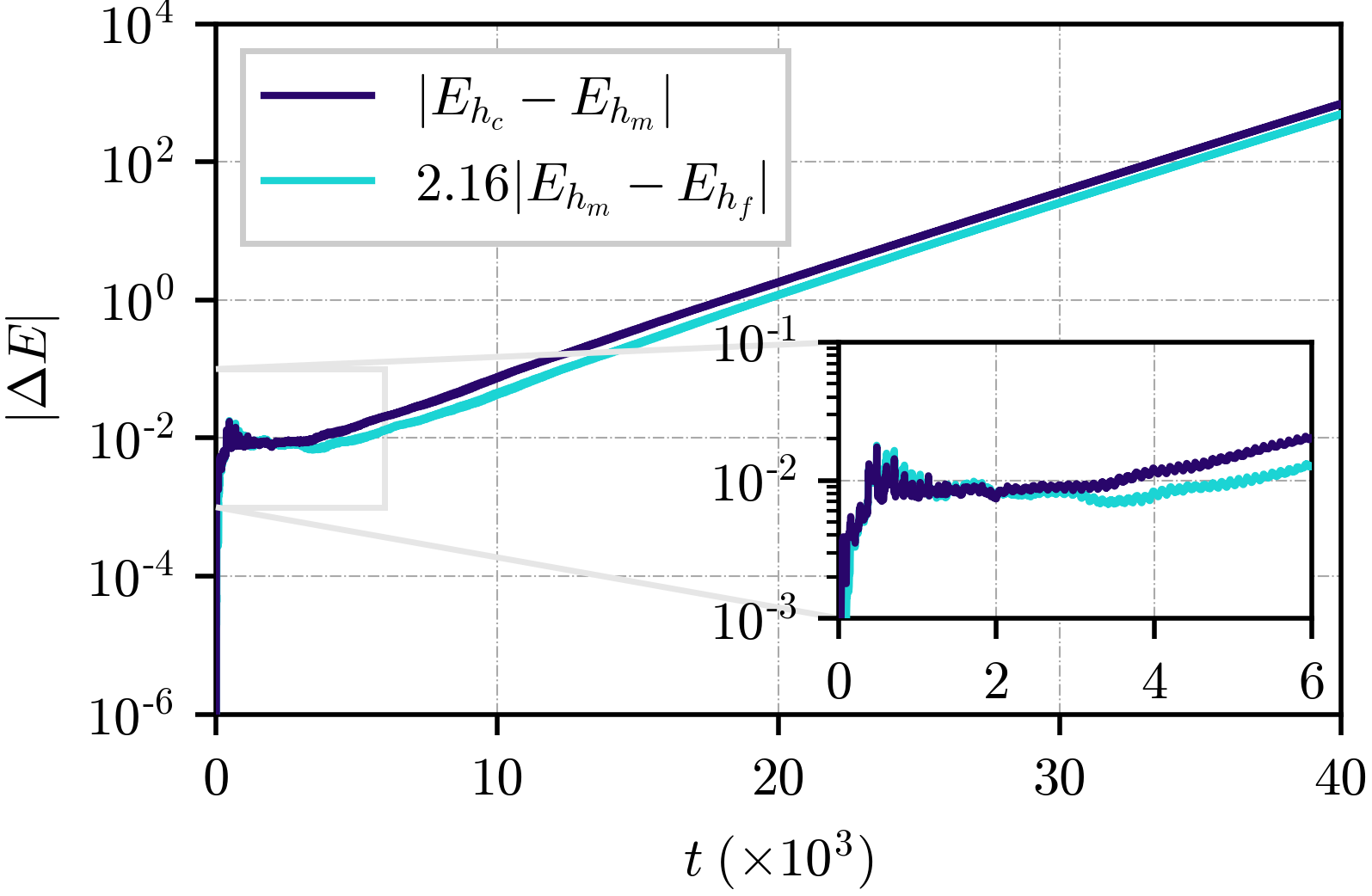}
\hfill
\includegraphics[width=0.475\linewidth]{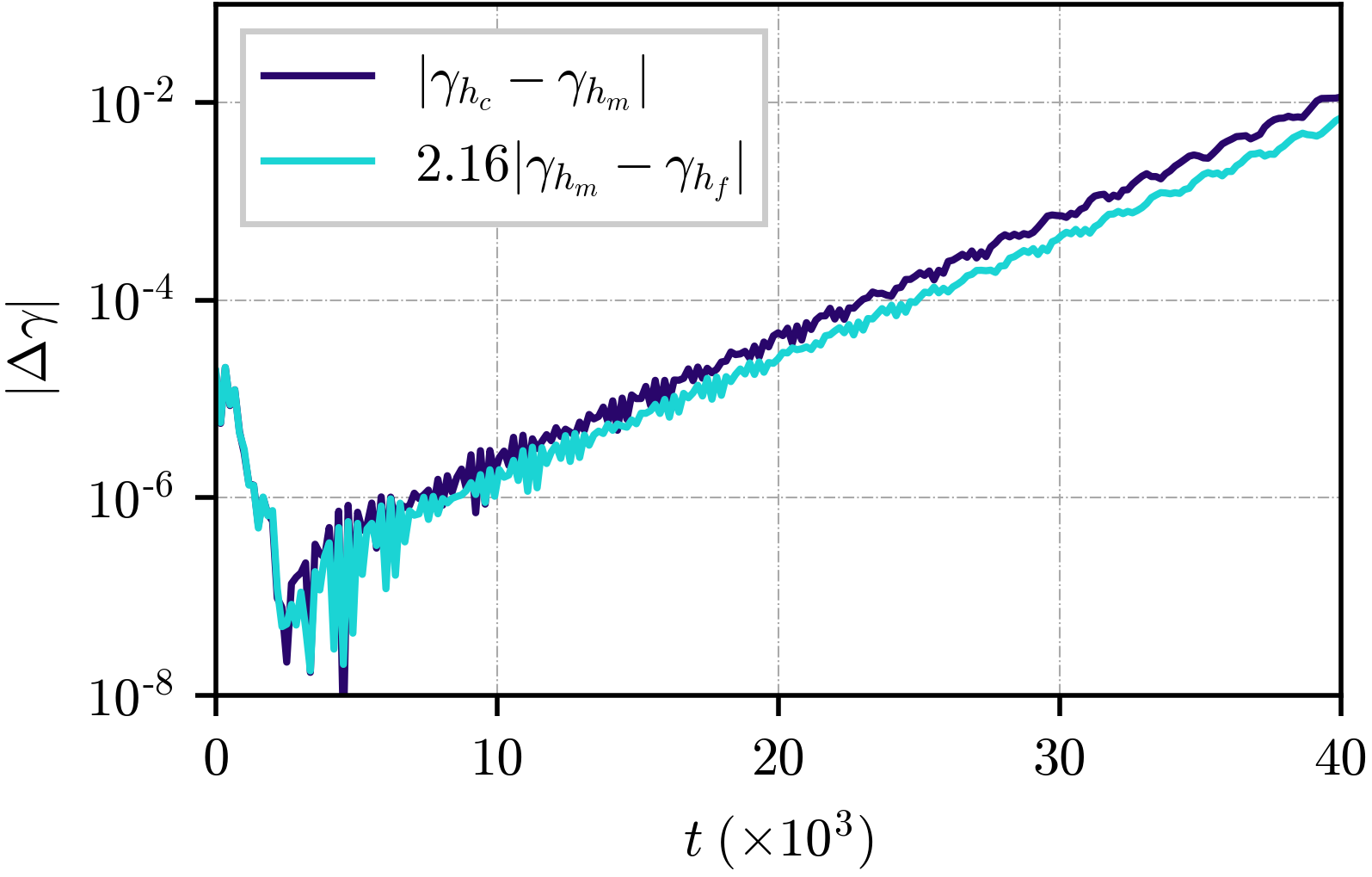}
\caption{Convergence analysis of the simulation shown in Fig.~\ref{fig:binary_bh_mural} in the main text, for the energy density inside the cavity (left panel) and average energy growth rate inside the cavity (right panel). Results are consistent with 2nd order convergence. 
\label{fig:energy_convergence}
}
\end{figure*}

The finite difference methods employed for the simulations
should approximate the continuum solution of the problem with an error that depends polynomially on the grid spacing~$h$,
\begin{equation}
f = f_h + \mathcal{O}(h^n) \; ,
\end{equation}
where $n$ is the convergence order. Since we use 2nd-order accurate operators,
we expect to see 2nd order convergence. This can be easily tested by running the same configuration for three different resolutions and calculating the $Q$-factor
\begin{equation}
	Q = \frac{h_c^n - h_m^n}{h_m^n - h_f^n} = \frac{f_{h_c}-f_{h_m}}{f_{h_m}-f_{h_f}} \; , 
	\label{eq:Q-factor}
\end{equation}
where $h_c$, $h_m$ and $h_f$ refer respectively to coarse, medium and fine grid resolutions.
We ran the configuration presented in Fig.~\ref{fig:binary_bh_mural} with the resolutions $h_c = 0.4013M$, $h_m = 0.3008M$ and $h_f = 0.24048M$
% (corresponding to 300, 400 and 500 nodes respectively)
and evaluated the energy content of the field inside the cavity at each iteration. For this set of resolutions the expected $Q$-factor for 2nd order convergence is $Q \simeq 2.16$. The results are summarized in Fig.~\ref{fig:energy_convergence}, and are consistent with 2nd-order convergence.
%

%%%%%%%%%%%%%%%%%%%%%%%%%%%%%%%%%%%%%%%%%%%%%%%%%%%%%%%%%%%%%%%%%%%%%%%%%%%%%%%%%%%%%%%
\section{Superradiance with connected bodies\label{appendix:appendix_cylinder_model}	}
%%%%%%%%%%%%%%%%%%%%%%%%%%%%%%%%%%%%%%%%%%%%%%%%%%%%%%%%%%%%%%%%%%%%%%%%%%%%%%%%%%%%%%%

\begin{figure}[ht]
\centering
\includegraphics[width=0.5\linewidth]{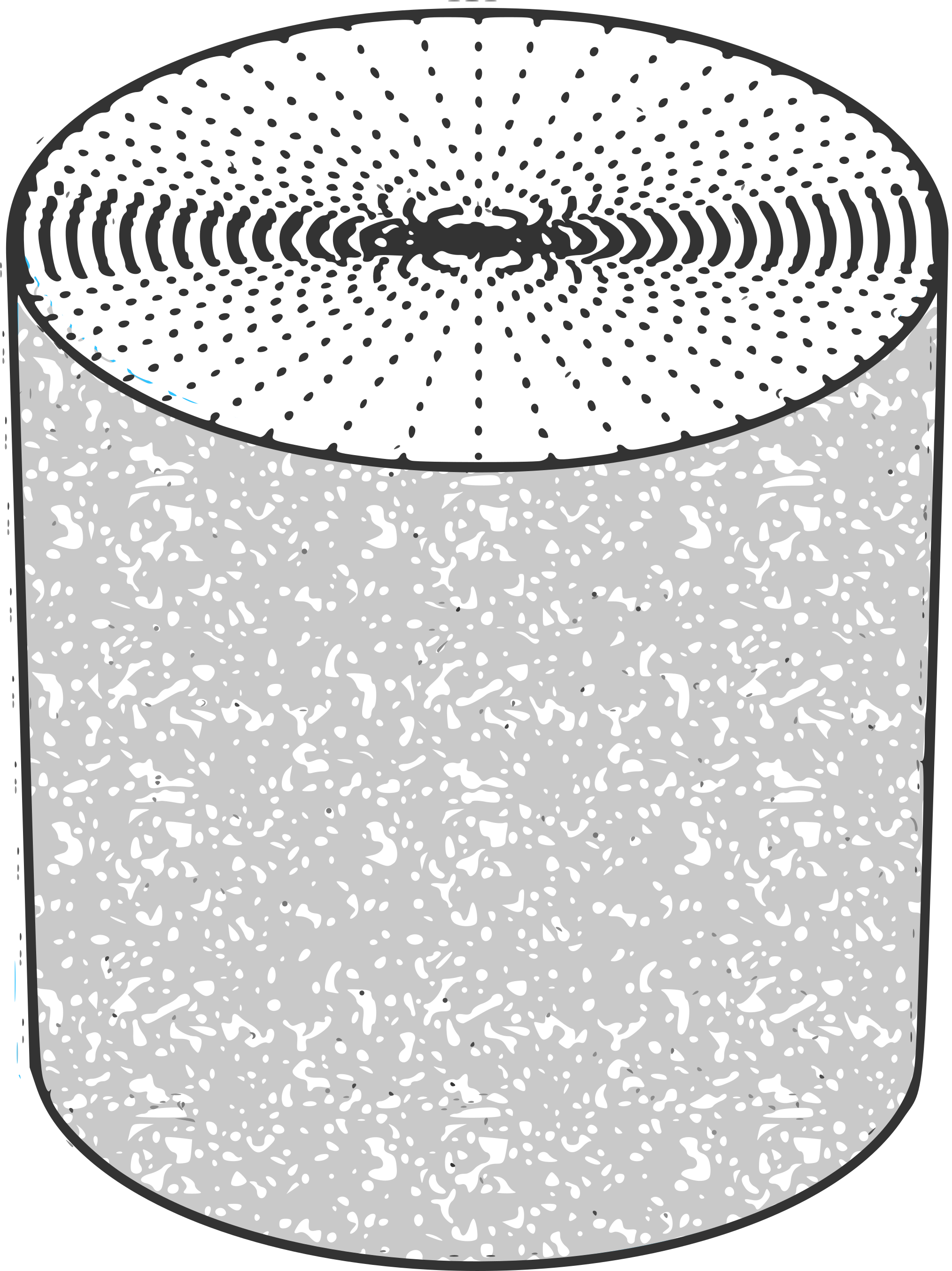}
\caption{Cartoon of a cylinder made of a inhomogeneous material. The dotted area on the surface depicts the magnitude of the impedance $Z$, which varies with azimuthal angle $\varphi$ along the geometrical axis of symmetry. For the example worked out numerically $Z = Z_0 (1 + \epsilon \cos 2 \varphi)$, mimicking the symmetries in the BHB problem.	
\label{fig:cylinder_model}
}
\end{figure}

The higher harmonic generation observed in Sec.~\ref{sec:binary_bh} is also present in other asymmetric setups. A very simple system (perhaps of physical relevance as well) which mimics the asymmetric geometry of a BHB is shown in Fig.~\ref{fig:cylinder_model}. It is a cylinder made of a certain material whose absorption properties vary with azimuth angle. 

Consider sound waves in a fluid where such cylinder is immersed. These are governed exactly by the same equations of our model. Namely, if the fluid perturbations are described by the scalar field $\Psi$, their dynamics will be governed by the KG equation $\square \Psi = 0$. In polar coordinates $(t,r,\varphi)$, we can use the ansatz $\Psi = e^{-i \omega t + i m \varphi } \phi(r) / \sqrt{r} $ to obtain the equation
\begin{equation}
	\frac{\partial^{2} \phi}{\partial r^{2}}+\left(\omega^{2}-\frac{m^{2}}{r^{2}}+\frac{1}{4 r^{2}}\right) \phi=0 \; ,
	\label{eq:altered_bessel_1} \; .
\end{equation}
%ansatz $\Psi = \phi r^{-1/2}$and using the ansatz \eqref{eq:decomposition} we have the governing equation \mz{nesta secção a notação está confusa. usamos $\psi$, $\phi$, $\Psi$ e $\psi_m$! não pode ser, é preciso uniformizar isto e explicar as quantidades todas.}
%
The general solution of our field is thus written in terms of the Bessel functions of the first and second kind as
%whose solution can be written in terms of Bessel functions
%
\begin{equation}
	\Psi(t,r,\varphi) = \left[ A  J_m \left( \omega r \right) + B Y_m\left( \omega r \right) \right] e^{-i \omega t + i m \varphi} \; .
\end{equation}
The cylinder here considered has an impedance (measured in its rest frame) given by
\begin{equation}
	Z = Z_0 \left[1 + \varepsilon \cos^2 (\varphi)\right] \; ,
	\label{eq:impedance_dependence}
\end{equation}
where $\varepsilon$ is an dimensionless parameter that measures the asymmetry of the system.
%\mz{o $\epsilon$ já foi usado... se calhar usar aqui outra letra (e defini-la!)}
The impedance enters the boundary condition at the cylinder's surface
\begin{equation}
	\left(\frac{\partial \Psi}{\partial t}\right)=- \tilde{Z} \left(\frac{\partial \Psi}{\partial r}\right) \,,
	\label{eq:boundary_condition}
\end{equation}
where $\tilde{Z} = Z/\rho$. If we take the cylinder to be static, we can understand the effect of the asymmetry when scattering waves off the cylinder. In this scenario, the solutions are better expressed as
\begin{equation}
	\Psi^{m}(t, r, \varphi)=\left[\mathcal{A}_{+}^m \phi_{+}^{m}\left(\omega r\right)+\mathcal{A}_{-}^m \phi_{-}^{m}\left(\omega r\right)\right] e^{-i \omega t+i m \varphi} \; ,
	\label{eq:hankel_decomposition}
\end{equation}
where $\phi_+^m$ and $\phi_-^m$ represent the Hankel functions of the first and second kind respectively.

After some algebra, using Eq.~\eqref{eq:boundary_condition}, the incoming and outgoing coefficients of \eqref{eq:hankel_decomposition} can be obtained. The expression is rather messy but can be encapsulated in the matrix equation
\begin{equation}
	\mathbf{M}_{+} \psi_{+}+\mathbf{M}_{-} \psi_{-}=0
\end{equation}
where the matrices and vectors are defined as
\begin{equation}
	\mathbf{M}_{\pm} \psi_{\pm}
	=
	\left[
	\begin{array}{ccccl}
		\ddots          & \beta_{\pm}^{2}   &                   &                   &                 \\
		\beta_{\pm}^{4} & \Lambda_{\pm}^{2} & \beta_{\pm}^{0}   &                   &                 \\
		& \beta_{\pm}^{2}   & \Lambda_{\pm}^{0} & \beta_{\pm}^{2}   &                 \\
		&                   & \beta_{\pm}^{0}   & \Lambda_{\pm}^{2} & \beta_{\pm}^{4} \\
		&                   &                   & \beta_{\pm}^{2}   & \ddots
	\end{array} 
	\right]
	\left[
	\begin{array}{c}
		\vdots \\
		\mathcal{A}_{\pm}^{-2} \\
		\mathcal{A}_{\pm}^{0} \\
		\mathcal{A}_{\pm}^{2} \\
		\vdots
	\end{array}\right]
\end{equation}
with coefficients given by
\begin{equation}
	\begin{gathered}
		\Lambda_{\pm}^{m}=\left[\phi_{\pm}^{m}-i \tilde{Z}\left(\phi_{\pm}^{m}\right)^{\prime}\right]+\epsilon\left(\frac{i \tilde{Z}}{2}\right)\left(\phi_{\pm}^{m}\right)^{\prime} \; , \\[10pt]
		\beta_{\pm}^{m}=\epsilon\left(\frac{i \tilde{Z}}{2}\right)\left(\phi_{\pm}^{m}\right)^{\prime} \; .
	\end{gathered} 
	\label{eq:matrix_components}
\end{equation}
In the above expressions, all functions should be evaluated at the cylinder's radius and the primes denote derivatives with respect to $y = \omega r/c$.

\begin{figure}[ht]
\centering
\includegraphics[width=\linewidth]{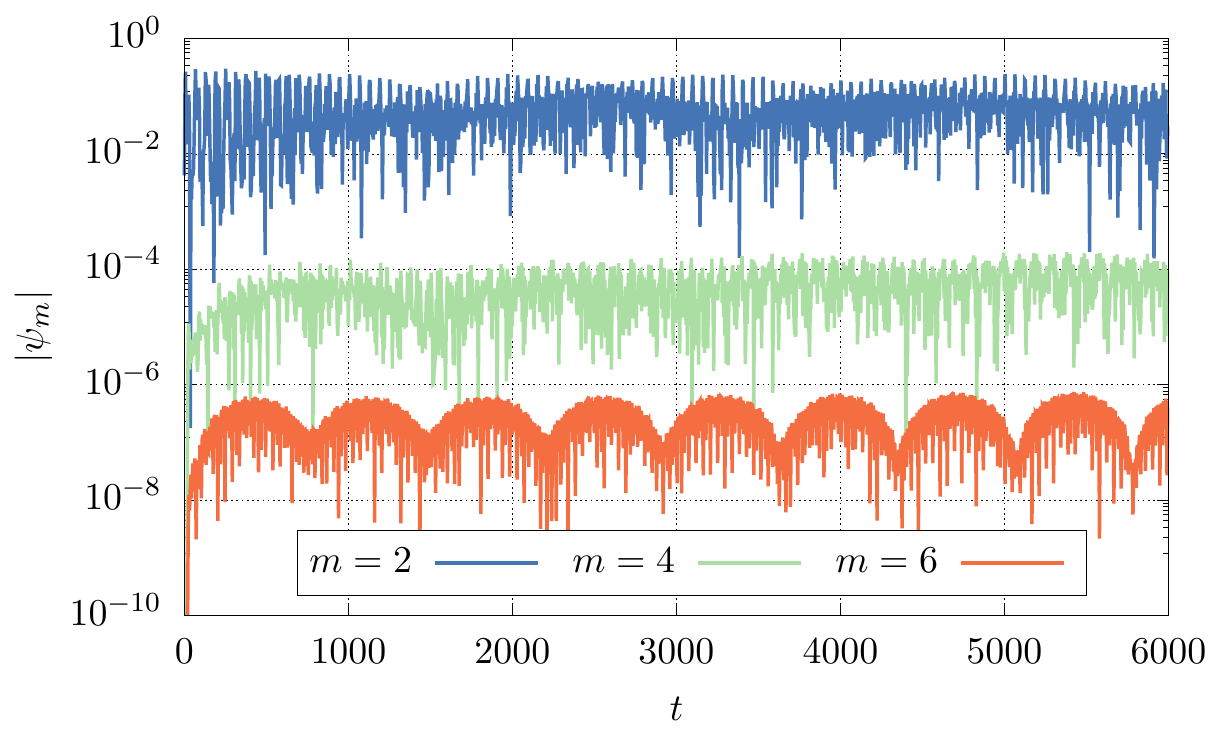}
\caption{Fourier components of an initial Gauss-like wave-packet~\eqref{eq:initial_pulse} with $\sigma=2$, $\omega=0.4$ and $r_0=15$ evaluated $r=10$. The inner cylinder has radius equal to unity, angular velocity $\Omega = 0.4$ and impedance $Z_0 = 10^{-3}$ while the reflective boundary is at $r = 31$. \label{fig:mode_mixing}
}
\end{figure}
The most prominent feature of this calculation is that the asymmetry only couples equal parity modes. 
This feature is also observed when the cylinder is considered to be rotating. Despite not being able to relate the coefficients analytically, a numerical evolution of the KG equation with the appropriate boundary conditions shows that in fact, for an initially $m=0$ Gaussian pulse, only even modes get excited (see Fig.~\ref{fig:mode_mixing}).
Note that setting $\epsilon = 0$ we recover the results obtained in~\cite{Cardoso:2016zvz} for a static uniform cylinder.

\bibliography{./bib/bibliography}

\end{document}